\documentclass[journal]{IEEEtran}
\ifCLASSOPTIONcompsoc
  \usepackage[nocompress]{cite}
\else
  \usepackage{cite}
\fi
\usepackage[hidelinks]{hyperref}
\usepackage{amsmath}
\ifCLASSINFOpdf
  \usepackage[pdftex]{graphicx}
\else
  \usepackage[dvips]{graphicx}
\fi

\usepackage{subcaption}
\usepackage[linesnumbered,ruled]{algorithm2e}

\SetAlFnt{\small}
\SetAlCapFnt{\small}
\SetAlCapNameFnt{\small}
\SetAlCapHSkip{0pt}
\IncMargin{-\parindent}

\usepackage{graphicx}
\usepackage{amssymb}
\usepackage{amsmath,url}
\usepackage{multicol,multirow}
\usepackage{epsf,psfrag,pstricks}
\usepackage{rotating}
\usepackage{multirow}
\usepackage{float}
\usepackage{bm}
\usepackage{pdflscape}
\usepackage[normalem]{ulem}
\usepackage{booktabs}
\usepackage{multirow}

\usepackage{array}
\newcolumntype{L}[1]{>{\raggedright\let\newline\\\arraybackslash\hspace{0pt}}m{#1}}

\newcommand{\eg}{{\it e.g.}}
\newcommand{\ie}{{\it i.e.}}

\newtheorem{definition}{Definition}
\newtheorem{theorem}{Theorem}

\begin{document}
\title{HyObscure: Hybrid Obscuring for Privacy-Preserving Data Publishing 
}

\author{
        Xiao~Han,
        Yuncong~Yang,
        Junjie~Wu
\thanks{Manuscript received xxx; revised xxx.}
\thanks{Xiao Han (e-mail: xiaohan@mail.shufe.edu.cn) and Yuncong Yang (e-mail: yycphd@163.sufe.edu.cn) are with Shanghai University of Finance and Economics, 777 Guoding Road, Shanghai 200433, China PR. Junjie Wu (e-mail: wujj@buaa.edu.cn) is with Beihang University, 37 Xueyuan Road, Beijing, 100191, China PR. Junjie Wu is the corresponding author.}
}

\markboth{IEEE Transactions on Dependable and Secure Computing}
{}

\maketitle
\IEEEpeerreviewmaketitle
\begin{abstract}
Minimizing privacy leakage while ensuring data utility is a critical problem to data holders in a privacy-preserving data publishing task. Most prior research concerns only with one type of data and resorts to a single obscuring method, \eg, obfuscation or generalization, to achieve a privacy-utility tradeoff, which is inadequate for protecting real-life heterogeneous data and is hard to defend ever-growing machine learning based inference attacks. This work takes a pilot study on privacy-preserving data publishing when both generalization and obfuscation operations are employed for heterogeneous data protection. To this end, we first propose novel measures for privacy and utility quantification and formulate the hybrid privacy-preserving data obscuring problem to account for the joint effect of generalization and obfuscation. We then design a novel hybrid protection mechanism called HyObscure, to cross-iteratively optimize the generalization and obfuscation operations for maximum privacy protection under a certain utility guarantee. The convergence of the iterative process and the privacy leakage bound of HyObscure are also provided in theory. Extensive experiments demonstrate that HyObscure significantly outperforms a variety of state-of-the-art baseline methods when facing various inference attacks under different scenarios. HyObscure also scales linearly to the data size and behaves robustly with varying key parameters. 

\end{abstract}


\maketitle

\section{Introduction}
In the era of big data, data publishing has become a popular way to facilitate data exploitation and enlarge economic value for data holders~\cite{Jia2017AttriInferIU,yang2019privacy,salamatian2015managing}. Many leading data holders like Facebook and Twitter provide API to share data with third-parties and increase platform engagements ~\cite{yang2019privacy}. Data holders are also more and more willing to publish data samples, such as \textit{Mimic database}\footnote{https://mimic.mit.edu/docs/iv/}, \textit{MovieLens}\footnote{https://netflixprize.com/}, and \textit{Yelp challenges}\footnote{https://www.yelp.com/dataset/challenge}, 
to seek worldwide help in data exploitation. As positive externality, data publishing enables service innovation, scientific discovery, and other public benefits, which generate enormous economic value amounting to over \$3 trillion annually\footnote{https://www.mckinsey.com/business-functions/mckinsey-digital/our-insights/open-data-unlocking-innovation-and-performance-with-liquid-information}.

\begin{figure}[t!]
	\centering
    \footnotesize
	\begin{subfigure}[t]{\linewidth}
		\includegraphics[width=\linewidth]{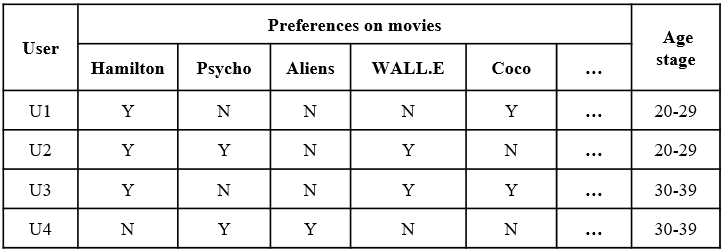}
		\caption{\emph{MovieLens} dataset. The preferences of users on every movie (privacy-insensitive data) and users' age stage (generalized private data) are published.}
		\label{fig:neflixData}
	\end{subfigure}
	\\
	\begin{subfigure}[t]{\linewidth}
		\includegraphics[width=\linewidth]{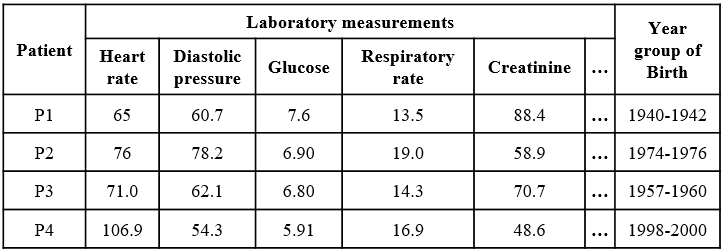}
		\caption{\emph{Mimic\_IV} database. It publishes patients' laboratory measurements (privacy-insensitive data) and the range of their birth year (generalized private data).}
		\label{fig:mimicData}
	\end{subfigure}
	\caption{Two public datasets containing both generalized private data and privacy-insensitive data.}\label{fig:dataExm}
	\vspace{-1.5em}
\end{figure}

While data publishing creates substantial benefits, privacy leakage due to improper sharing of user data may get data holders into great troubles, even the situation involving multi-billion dollar fines or lawsuits\footnote{https://www.forbes.com/sites/mnunez/2019/07/24/ftcs-unprecedented-slap-fines-facebook-5-billion-forces-new-privacy-controls/}. As a result, it is a critical task for data holders to protect users' privacy when exploring sustainable data benefits. In practice, data holders typically restrict the access to users' \emph{exact private} data that users are reluctant to share (\eg, age or home address) for privacy protection, and release \emph{privacy-insensitive} information (\eg, movie ratings) that users agree to open in return for favorable services~\cite{Alessandro2015Privacy}. Nevertheless, \emph{generalized private} data (\eg, age stage or living city) are often publicly available in many datasets~\cite{MovieLens1M,ChicagoTaxi}, since these data cause few direct privacy breaches but are required by or helpful to services like health management or personalized recommendation. 
Fig.~\ref{fig:dataExm} visualizes a few data samples of two public data sets, in which both privacy-insensitive data and generalized private data are present.

With the unprecedented development of artificial intelligence technology, however, the above-mentioned data publishing routine may not truly protect users' privacy, especially with the recently emergent \emph{attribute inference attacks}~\cite{gong2018attribute,chaabane2012you,dey2012estimating,han2015alike,heatherly2012preventing}. 
For instance, users’ hidden attributes such as ages, political views and locations are likely to be predicted by users' publicly available data such as their ``likes'', posts, and relationships in online social media~\cite{Kosinski2013PrivateTA,FPAD2019}, which could invalidate the generalization of private data. 
In this light, how to achieve \emph{privacy-preserving data publishing}, so that users' exact private data can be highly protected from attribute inference attacks while useful privacy-insensitive data are published, has become an open challenge to both academia and industry.

Delicately obscuring data to trade minimum data utility for maximum privacy protection is a general idea to accomplish privacy-preserving data publishing against attribute inference attacks~\cite{Sankar2013UtilityPrivacyTI}. A few recent studies focus on privacy-insensitive data to devise optimal obscuring (\eg, obfuscation~\cite{yang2019privacy}
or noise addition~\cite{AttriGuard}) methods. These methods, however, will encounter problems when it requires to publish both privacy-insensitive and generalized private data. Specifically, how to generalize the exact private data for publishing needs a design, and more importantly, the correlation between privacy-insensitive and generalized private data may require synthetic obscuring methods to attain a joint effect of privacy protection. Some other studies obscure privacy sensitive and insensitive data together by generalization methods~\cite{Microaggregation2002}. They typically group users with similar attributes and represent all the users in a group with the same generalized attribute values for data publishing. Nevertheless, suffering from the curse of dimensionality, leveraging generalization methods only tends to be impractical when datasets contain a large number of attributes~\cite{aggarwal2005on}. 

In light of the above difficulties in utilizing one obscuring operation on real-life heterogeneous data for privacy protection, it is a natural consideration to apply diverse operations, \eg, generalization and obfuscation, to different parts of data where both privacy-insensitive and generalized private data are present. For instance, generalization could be applied only to private attributes of a small set so that the operation is efficient and no exact private information is published. 
Obfuscation, commonly regarded as a feasible obscuring operation to high-dimensional data~\cite{Yang2016PrivCheckPC}, could be applied on privacy-insensitive data to ensure fine-grained data publishing, which are often deemed more valuable. For instance, the ratings on specific movies are more valuable than that on movie categories (\eg, drama). Along this line, in this paper, we propose a hybrid obscuring solution that obfuscates privacy-insensitive data and generalizes private data synthetically for favorable privacy-preserving data publishing. 

There are two major challenges for hybrid obscuring. The first one is to quantify the data utility loss and the privacy gain so that we can find a good tradeoff. Existing quantification methods are designed for a single obscuring operation~\cite{Jia2017AttriInferIU,Yang2016PrivCheckPC}, which could be employed separately but would neglect the potential interactive impacts of hybrid obscuring operations stemming from intrinsic data correlations. The other challenge comes from identifying the obfuscation and generalization functions that can jointly optimize the privacy-utility tradeoff. We should deal with two interleaved sets of variables for optimizing hybrid obscuring, whereas prior work typically optimizes a solvable objective function with one set of variables for a single obscuring
approach~\cite{Jia2017AttriInferIU,yang2019privacy}. In general, this work overcomes the above challenges and makes the following contributions:

1) It formulates a hybrid privacy-preserving data obscuring problem, which aims to minimize the privacy leakage under a guaranteed data utility loss with two delicately designed obscuring functions for users' privacy-insensitive and private data, respectively. To the best of our knowledge, this is the first work to investigate hybrid obscuring techniques for privacy-preserving data publishing.

2) It proposes a new method to quantify privacy leakage caused by the release of hybrid obfuscated privacy-insensitive and generalized private data. It can also deal with data utility loss computations given different data obscuring techniques.

3) It decomposes the complicated optimization problem into three sub-problems, namely $I$-problem (Initialization), $O$-problem (Obfuscation), and $G$-problem (Generalization), and proposes a hybrid obscuring approach called \emph{HyObscure} to cross-iteratively optimize the privacy-utility tradeoff. Theoretically proved convergence and privacy leakage bounds are also provided to HyObscure.

We conduct extensive experiments on two real-life data sets, with the challenges of two inference attack methods and under two attacker prior knowledge scenarios, to validate the effectiveness of HyObscure. Results demonstrate that HyObscure provides consistently much better privacy-utility tradeoffs compared with state-of-the-art baselines, scales linearly to data size, and is robust to parametric settings. The remainder of this work is organized as follows. We formulate our hybrid obscuring problem in Sect.~\ref{sec:problem}, and introduce the HyObscure algorithm and its experimental performances in Sect.~\ref{sec:hyobscure} and Sect.~\ref{sec:experiment}, respectively. We present the related work in Sect.~\ref{sec:relate} and finally conclude our work in Sect.~\ref{sec:conclude}. We leave detailed proofs to theorems to Supplemental Document for concise presentation.
\section{Problem Formulation}
\label{sec:problem}

\begin{table}[t!]
	\centering
	\footnotesize
	\renewcommand*{\arraystretch}{1.2}
	\caption{Summary of math notations.}
	\vspace{-0.5em}
	\begin{tabular}{cl}
		\toprule
		Notation       & \multicolumn{1}{c}{Description} \\
		\midrule
		$X$ ($\hat{X}$)              & Original (obfuscated) insensitive data.      \\
		$Y$ ($\widetilde Y$)            & Original (generalized) private data.       \\
		$\mathcal{O}_{\hat X|X}$   & Obfuscation function.  \\
		$\mathcal G_{\widetilde Y|Y}$    & Generalization function. \\
		$\mathcal{G}^0_{\widetilde Y|Y}$     & Initial generalization function. \\
		$y$ ($\tilde y$)     & The exact (generalized) value of private data. \\
		$X_{\tilde y}$ ($\hat X_{\tilde y}$) & Original (obfuscated) insensitive data of users with $\tilde y$.\\
		$Y_{\tilde y}$ & The exact values of private data associated with $\tilde y$. \\
		$I(\cdot)$      & Mutual information. \\
		\bottomrule
	\end{tabular}
	\label{tbl:notations}
	\vspace{-1.5em}
\end{table}

In this work, we address a hybrid privacy-preserving data publishing problem where both users' privacy-insensitive data and exact private data are given. Basically, by designing an obfuscation function $\mathcal{O}_{\hat{X}|X}$ for users' privacy-insensitive data $X$ and a generalization function $\mathcal{G}_{\widetilde Y|Y}$ for users' private data $Y$, we expect to attain two objectives: 1) to reserve the data utility for downstream services, and 2) to protect the exact private data when the obfuscated privacy-insensitive data $\hat{X}$ and the generalized private data $\widetilde Y$ are published concurrently. In what follows, we first quantify data utility and privacy leakage, and then formulate our problem. Table ~\ref{tbl:notations} summarizes the notations used throughout this paper.


\subsection{Obfuscation and Generalization}
Obfuscation and Generalization are two widely used data obscuring techniques. In general, obfuscation does not change much fineness of data while generalization makes data coarse.

\textbf{Obfuscation}.~By carefully designing a probabilistic obfuscation function $\mathcal{O}_{\hat{X}|X}$ given an obfuscation budget, obfuscation techniques switch users' privacy-insensitive data $X$ by other users' data $\hat{X}$ with certain probabilities, in order to reduce the disclosure probability of users' private information under a guarantee of data utility~\cite{salamatian2015managing,yang2019privacy}.

\textbf{Generalization}.~A generalization function~$\mathcal G_{Y|\widetilde Y}$  maps an arbitrary exact value of a private attribute $y \in Y$ into a generalized value $\tilde y \in \widetilde Y$. Each generalized value $\tilde y \in {\widetilde Y}$ associates with a group of exact values, denoted as $Y_{\tilde y}$. For instance, the exact age of $38$ can be generalized to an age stage of $30$--$40$. The obscured groups of exact private values are usually required to satisfy users' privacy requirements such as \textit{K}-anonymity \cite{Sweeney2002kAnonymityAM} and $L$-diversity~\cite{Machanavajjhala2006lDiversityPB}.

\textbf{Hybrid Obfuscation and Generalization}.~Given users' privacy-insensitive data $X$ and private data $Y$, $X$ is obfuscated and $Y$ is generalized concurrently for publishing, so that the exact value of $Y$ is preserved and the fineness of $X$ is maintained. In particular, to keep the privacy-insensitive value distribution $X_{\tilde y}$ invariant after obfuscation for $\tilde y \in \widetilde Y$, the obfuscation of $X$ is only allowed between users with the same $\tilde y \in \widetilde Y$~\cite{willenborg2012elements}.



\subsection{Privacy Leakage Computation}
If the obfuscated privacy-insensitive data $\hat{X}$ and generalized private data $\widetilde Y$ are published, attackers may perform attribute inference attacks to correctly infer users' private data $Y$ based on the published data, which will incur privacy leakage on $Y$.

\begin{definition}[Attribute Inference Attack]
	Given $[\hat{X}, \widetilde Y]$ as the published data of a set of users $U$, an attribute inference attack is to infer users' exact values of $Y$ with the objective of minimizing the expected inference loss:
	\begin{equation}
	q^* = \arg \min_{q} E_{_{Y|\hat{X}, \widetilde Y}} [\mathcal{L}(Y,q)|\hat{X}, \widetilde Y],
	\end{equation}
	where $\mathcal{L}$ denotes the loss function
	, and $q^*$ is the preferable inference method obtained by solving the attribute inference attack problem.~\hfill$\blacksquare$
\end{definition}

Given the above definition, the privacy leakage on $Y$ for publishing $[\hat{X}, \widetilde Y]$ can then be defined as the inference gain (\ie, the decrease of inference loss) of attackers owing to procuring $[\hat{X}, \widetilde Y]$:
\begin{equation}\label{eq3}
\Delta \text{loss}_\text{privacy} = \min_{q} E_{_{Y|\hat{X}, \widetilde Y}} [\mathcal{L}(Y,q)|\hat{X}, \widetilde Y] - \min_{q} E_{_{Y}} [\mathcal{L}(Y,q)].
\end{equation}

We adopt the \textit{log-loss} function as the loss function $\mathcal{L}$ in Eq.~\eqref{eq3}, which has been widely used in general inference tasks and is deemed consistent with ``rational'' adversaries' goals~\cite{Calmon2012PrivacyAS}. We also highlight in Sect.~\ref{Sec2_Discussion} that the privacy leakage computed with the log-loss function relates to the error ratio of broad adversary inference algorithms. Specifically, before observing $\hat{X}$ and $\widetilde{Y}$, the expected loss of adversaries using log-loss can be rewritten as:
\begin{equation}
\begin{aligned}
E_{_{Y}} [\mathcal{L}(Y,q)] &= \sum_{y\in Y} -{p(y)\log q}\\
&\ge \sum_{y\in Y} -{p(y)\log p(y)} = H(Y).
\end{aligned}
\label{eq4}
\end{equation}
After observing $\hat{X}$ and $\widetilde{Y}$, the expected loss turns to:
\begin{equation}
\begin{aligned}
E_{_{Y|\hat{X}, \widetilde Y}} [\mathcal{L}(Y,q)|\hat{X}, \widetilde Y] &= \sum_{y\in Y} -{p(y|\hat{X},\widetilde{Y})\log q(\hat{X},\widetilde{Y})} \\ 
&\ge \sum_{y\in Y} -{p(y|\hat{X},\widetilde{Y})\log(p(y|\hat{X},\widetilde{Y}))} \\
&= H(Y|\hat{X},\widetilde{Y}).
\end{aligned}
\label{eq5}
\end{equation}

According to Eqs.~\eqref{eq3} - \eqref{eq5}, we finally have:
\begin{equation}
\begin{aligned}
\Delta \text{loss}_\text{privacy} = H(Y|\hat{X},\widetilde{Y}) - H(Y) = I(Y;\hat{X},\widetilde{Y}).
\end{aligned}
\label{eq6}
\end{equation}
In brief,  when the log-loss function is used, $\Delta \text{loss}_\text{privacy}$ equals the mutual information between the published data $[\hat{X}, \widetilde Y]$ and the exact values of private attributes $Y$, which can be computed following Theorem~\ref{theo:leakage} below (we leave proofs to the supplemental document for concision)

\begin{theorem}\label{theo:leakage}
	The privacy leakage on $Y$ caused by the publicly released $[\hat{X}, \widetilde Y]$ is a function of $\mathcal{O}_{\hat X|X}$ and $\mathcal{G}_{\widetilde{Y}|Y}$, which is given by
	\begin{equation}
	\begin{aligned}
	&\Delta \text{loss}_\text{privacy} = I(Y;\hat X,\widetilde{Y}) \nonumber\\
	\end{aligned}
	\end{equation}
	\begin{equation}\label{eq:finalMutualInfo}
	\footnotesize
	\begin{aligned}
	& =\sum_{\substack{\tilde y \in \widetilde Y\\ y' \in Y}} \mathcal{G}_{\widetilde{Y}|Y}(\tilde y|y') p_Y(y')  \sum_{\substack{\hat x\in \hat X_{\tilde y}\\ x\in X_{\tilde y}\\y \in Y_{\tilde y}}} \mathcal{O}_{\hat X|X}(\hat x|x, \tilde y) \ p_{XY}(x,y|\tilde y) \\
	& \log \frac{\sum \limits_{x'\in X_{\tilde y}} \mathcal{O}_{\hat X|X}(\hat x|x', \tilde y) \ p_{XY}(x',y| \tilde y)}{ \sum_{\substack{y'\in Y_{\tilde y}\\x''\in X_{\tilde y}}} \mathcal{O}_{\hat X|X}(\hat x|x'', \tilde y) \  p_{XY}(x'',y'| \tilde y)}- \sum_{y \in Y} p_Y(y)\log p_Y(y),
	\end{aligned}
	\end{equation}
	where $p_{Y}(y')$, $p_{XY}(x,y|\tilde y)$ and $ \sum \limits_{y \in Y} p_Y(y)\log p_Y(y)$ can be calculated by specific data given $\mathcal{O}_{\hat X|X}$ and $\mathcal{G}_{\widetilde{Y}|Y}$.~\hfill$\blacksquare$
\end{theorem}
 

\subsection{Data Utility Maintenance}
\label{sub:qos}
Obfuscation on privacy-insensitive data $X$ may diminish the utility of~$X$, while generalization on private data $Y$ often lowers the utility of~$Y$. Data utility maintenance thus becomes an associated and important problem of privacy protection.


\subsubsection{Maintaining Utility for Obfuscation} Data utility loss caused by obfuscation is typically measured by the expected difference between the obfuscated and original data,\ie, $\mathbb{E}_{X\hat{X}}(d(X,\hat{X}))$,
where $d$ is a difference calibration function, such as $l_2$-norm, Kullback-Leibler divergence~\cite{klDivergence}, and Jensen-Shannon divergence~\cite{lin1991divergence}. 

In our case of hybrid obfuscation and generalization, as the obfuscation of $X$ is only allowed between users with the same generalized private value $\tilde y \in \widetilde Y$, the utility loss of obfuscation is computed by the expected distance between $X$ and $\hat X$ given $\widetilde Y$, which is given by
\begin{equation}\label{eq:Xquality}
\begin{aligned}
\mathbb{E}_{X\hat X}[d(X,\hat X)]  = 
  \mathbb{E}_{\tilde y \in \widetilde Y}\mathbb{E}_{X_{\tilde y}\hat X_{\tilde y}}[d(X_{\tilde y}, \hat X_{\tilde y})].
\end{aligned}
\end{equation}
In this way, we can control the data utility loss while applying generalization by the privacy requirements.





\subsubsection{Maintaining Utility for Generalization} Concerning that data utility loss due to generalization closely relates to the degree of generalization, we define two measures for maintaining data utility against generalization as follows.


\begin{definition}[$(k,\alpha)$-uniqueness]
	It controls the number of users $|U_{\tilde y}|$ given a generalized private attribute value $\tilde y \in \widetilde Y$, and is achieved if $k \leq |U_{\tilde y}| \leq \alpha, \forall \tilde y\in  {\widetilde Y}$.~\hfill$\blacksquare$
\end{definition}


\begin{definition}[$(l,\beta)$-variety]
	It maintains the size of exact values $|Y_{\tilde y}|$ regarding a generalized private attribute value $\tilde y \in \widetilde Y$, and is achieved if $l \leq |Y_{\tilde y}| \leq \beta, \forall \tilde y \in {\widetilde Y}$.~\hfill$\blacksquare$
\end{definition}



In general, data utility decreases if more users or more exact private attribute values are clustered regarding a generalized value. As a result, $\alpha$ and $\beta$ respectively limit the numbers of users and exact private attribute values in a generalized group to retain data utility. However, the privacy is likely violated if the size of users or exact private values in a group is too small. Therefore, $k$ and $l$ are set to ensure enough users and exact private attribute values in a generalized group for privacy protection~\cite{Sweeney2002kAnonymityAM,Machanavajjhala2006lDiversityPB}. 

\subsection{Problem Definition}
Given the above definitions, we formulate our hybrid privacy-preserving data obscuring problem as follows:
{\small
\begin{align}
    \min \limits_{\mathcal{O}_{\hat X|X}, \mathcal{G}_{\widetilde Y|Y}} \ \ & I(Y;\hat{X},\widetilde{Y})   \label{prob:objFun} \\
    \text{s.t.}\ \ \ \ \ \   &  \mathbb{E}_{X \hat X}[d(X, \hat X)] \le \Delta Q_X \label{prob:Xquality},  \\
    &k \leq |U_{\tilde y}|\leq \alpha, \forall \tilde y \in \mathcal{\widetilde Y}  \label{prob:Kprivacy},\\
    &l \leq |Y_{\tilde y}|\leq \beta, \forall \tilde y \in \mathcal{\widetilde Y} \label{prob:Lprivacy}, \\
    &  \mathcal{O}_{\hat X|X}(\hat x|x) \in [0,1], \forall x,\hat x\in {X} \label{prob:probability1}, \\
    & \sum \limits_{\hat x}  \mathcal{O}_{\hat X|X}(\hat x|x)=1,\forall x\in {X} \label{prob:probability2}.
\end{align}}

In general, we aim to strike a balance between privacy protection and utility maintenance by taking the minimization of privacy loss due to data publishing as the objective function and taking data utility loss as the constraints of the optimization. The obfuscation function $\mathcal{O}_{\hat X|X}$ and the generalization function $\mathcal{G}_{\widetilde Y|Y}$ are two key optimum operators we seek for solving the constrained optimization problem. In more detail, Eq.~\eqref{prob:Xquality} restricts data utility loss in publishing $\hat X$ instead of $X$, where $\Delta Q_X$ denotes the obfuscation budget. Eq.~\eqref{prob:Kprivacy} and Eq.~\eqref{prob:Lprivacy} control the utility and meanwhile preserve the privacy of generalized private data $\widetilde Y$; that is, higher data utility will be retained with smaller $|U_{\tilde y}|$ and $|Y_{\tilde y}|$. Eq.~\eqref{prob:probability1} and Eq.~\eqref{prob:probability2} are two general probability constraints.

\subsection{Discussion}
\label{Sec2_Discussion}
While our problem assumes that adversaries apply log-loss function in their attack methods, in fact they may use any inference methods with alternative loss functions. Nevertheless, the utilization of log-loss function is sensible. On one hand, it is impossible to enumerate all the inference methods  as well as loss functions when quantifying the precise privacy leakage. On the other, prior studies have proved that the inference success of any method indeed relates to the mutual information measure~\cite{Calmon2012PrivacyAS}. In particular, by Fano's Inequality, the error probability $p(Y\not = q^*(\hat{X},\widetilde{Y}))$ of any inference algorithm~$q^*$ that infers $Y$ given data $[\hat{X}, \widetilde Y]$ is lower-bounded by
\begin{equation}
\begin{aligned}
p(Y\not = q^*(\hat{X},\widetilde{Y})) \ge \frac{H(Y)-I(Y;\hat{X},\widetilde{Y})-1}{\log|\mathcal{Y}|}.
\end{aligned}
\label{eq14}
\end{equation}
Note that in Eq.~\eqref{eq14} $H(Y)$ and $\log|\mathcal{Y}|$ are fixed given certain private information $Y$. In this light, the lower bound will increase as $I(Y;\hat{X},\widetilde{Y})$ decreases. In other words, while our problem aims at minimizing $I(Y;\hat{X},\widetilde{Y})$, it tends to raise the inference error by maximizing its lower bound no matter which inference methods are applied.
\section{Hybrid Obscuring Method}
\label{sec:hyobscure}
\subsection{Overview}
\begin{figure*}[t!]
	\centering
	\includegraphics[width=0.97\linewidth]{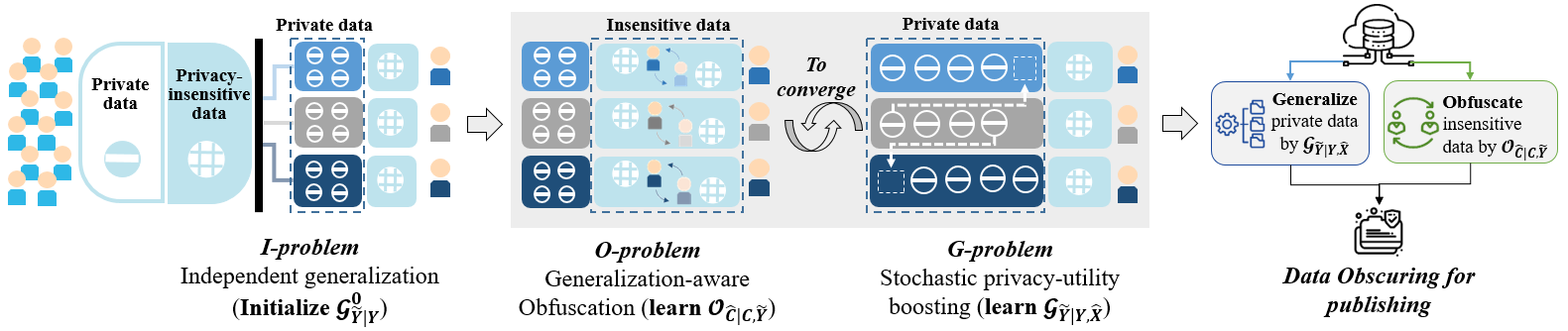}
	\caption{An overview of HyObscure.}
	\vspace{-1.5em}
	\label{fig:overview}
\end{figure*}

The obfuscation function $\mathcal{O}_{\hat X|X}$ and generalization function $\mathcal{G}_{\widetilde Y|Y}$ are two interleaved sets of decision variables in our objective function (see Eq.~\eqref{prob:objFun}). Since the objective function is not necessarily linear or convex, solving it directly is often intractable. We then resort to problem decomposition for a feasible solution. Specifically, we propose \textit{HyObscure} to systematically solve the problem by four steps: 1) decompose the hybrid privacy-preserving data obscuring problem into three sub-problems, \ie, $I$-problem (Initialization), $O$-problem (Obfuscation) and $G$-problem (Generalization); 2) solve $I$-problem to initialize $\mathcal G_{\widetilde Y|Y}$; 3) solve $O$-problem and $G$-problem in a cross-iterative manner until convergence to obtain $\mathcal O_{\hat X|X}$ and $\mathcal G_{\widetilde Y|Y}$; 4) conduct obfuscation and generalization on $X$ and $Y$ respectively by $\mathcal O_{\hat X|X}$ and $\mathcal G_{\widetilde Y|Y}$, and generate $\hat X$ and $\widetilde Y$ for publishing. Fig.~\ref{fig:overview} gives an overview of HyObscure.

\subsection{Problem Decomposition}

Decomposition technique is a general solution to optimization problems with complicated variables. The basic idea is to first break up the problem into a set of sub-problems and then separately solve each one of them. Intrinsically, the objective equation is convex regarding the obfuscation function $\mathcal{O}_{\hat X|X}$ when the generalized private data $\widetilde Y$ are certain; it is also solvable regarding the generalization function $\mathcal{G}_{\widetilde Y|Y}$ if the obfuscated privacy-insensitive data $\hat X$ are fixed. Along this line, we first carefully decompose the intractable problem into three solvable sub-problems, namely \textit{O-problem} for optimizing the obfuscation function $\mathcal{O}_{\hat X|X}$ given $\widetilde Y$, \textit{G-problem} for optimizing the generalization function $\mathcal{G}_{\widetilde Y|Y}$ given $\hat X$, and \textit{I-problem} for initiating the optimization. $O$-problem and $G$-problem are listed as follows:
{
\small
\begin{align}
     \textbf{O-problem:}\ \ \ \ \ \ \ &\min \limits_{\mathcal{O}_{\hat X|X, \widetilde Y}} \ \  I(\hat X, \widetilde Y; Y)   \label{eq:OsubobjFun} \\
    \text{s.t.}\ \ \ \ \ \   &  \mathbb{E}_{\tilde y\in
    	\widetilde Y} \mathbb{E}_{X \hat X}[d(X, \hat X| \tilde y)] \le \Delta Q_X \label{prob:OsubXquality},  \\
    &  \mathcal{O}_{\hat X|X}(\hat x|x) \in [0,1], \forall x, \hat x\in \mathcal{X} \label{eq:Osubprob1}, \\
    & \sum \limits_{\hat x}  \mathcal{O}_{\hat X|X}(\hat x|x)=1,\forall x\in \mathcal{X} \label{eq:Osubprob2},
\end{align}
}
{
\vspace{-1em}
\small
\begin{align}
    \textbf{G-problem:}\ \ \ \ \ \ \ &\min \limits_{\mathcal{G}_{\widetilde Y|Y, \hat X}} \ \ I(\hat X, \widetilde Y; Y)   \label{eq:GsubobjFun} \\
    \text{s.t.}\ \ \ \ \ \   &
     \mathbb{E}_{\tilde y\in
    	\widetilde Y} \mathbb{E}_{X \hat X}[d(X, \hat X| \tilde y)] \le \Delta Q_X,
    \label{prob:GsubXquality}
    \\
   &k \leq |U_{\tilde y}|\leq \alpha, \forall \tilde y \in \mathcal{\widetilde Y}  \label{eq:IsubKprivacy},\\
    &l \leq |Y_{\tilde y}|\leq \beta, \forall \tilde y \in \mathcal{\widetilde Y} \label{eq:GsubLprivacy},
\end{align}}
\noindent where $\mathcal{O}_{\hat X|X, \widetilde Y}$ is the obfuscation function given $\widetilde Y$, and  $\mathcal{G}_{\widetilde Y|Y, \hat X}$ is the generalization function given $\hat X$. 

When $\widetilde Y$ is fixed in $O$-problem, the privacy leakage  minimization objective function (see Eq.~\eqref{eq:OsubobjFun}) becomes convex  in terms of $\mathcal{O}_{\hat X|X, \widetilde Y}$, which can be efficiently solved by off-the-shelf optimization softwares such as CVX~\cite{CVX_2008}. Given a fixed $\hat X$, $\mathcal{G}_{\widetilde Y|Y, \hat{X}}$ is the only set of variables to be determined in $G$-problem. However, finding the optimal $\mathcal G_{\widetilde Y|Y, \hat X}$ is much harder as the objective function is not convex, for which a greedy algorithm will be designed.

To start the recursive optimization process of two decomposed sub-problems, the initial generalized private data or obfuscated public-insensitive data are needed. The independent generalization can be regarded as a clustering problem that clusters a set of nearby exact values $y$ into a generalized value $\tilde y$. So we formulate the $I$-problem as a typical clustering problem to obtain a good initial generalized private data $\widetilde Y^0$ as follows: 
{
\small
\begin{align}
    \textbf{I-problem:}\ \ \ \ \ \ \ &\min \limits_{\mathcal{G}^0_{\widetilde Y|Y}} \ \ \mathbb{E}_{\tilde y \in \widetilde Y} \mathbb{E}_{y\in Y_{\tilde y}}[d(y,c_{\tilde y})]  \label{eq:IsubobjFun} \\
    \text{s.t.}\ \ \ \ \ \   &k \leq |U_{\tilde y}|\leq \alpha, \forall \tilde y \in \mathcal{\widetilde Y}  \label{eq:IsubKprivacy},\\
    &l \leq |Y_{\tilde y}|\leq \beta, \forall \tilde y \in \mathcal{\widetilde Y} \label{eq:IsubLprivacy},
\end{align}
}
\noindent where the objective (Eq.~\eqref{eq:IsubobjFun}) is to minimize the expected sum of within-group distances between an arbitrary element inside each group $y\in Y_{\tilde y}$ and the centroid of the group $c_{\tilde y}$~\cite{Hartigan1979StatisticalAA}; Eq.~\eqref{eq:IsubKprivacy} and Eq.~\eqref{eq:IsubLprivacy} guarantee $(k,\alpha)$-uniqueness and $(l,\beta)$-variety, respectively. By solving the $I$-problem, the derived generalization function $\mathcal{G}^0_{\widetilde Y|Y}$ can be used to generate a good initial generalized private data $\widetilde Y^0$.

In brief, we decompose our hybrid data obscuring problem into three sub-problems. Next, we successively address these sub-problems.


\subsection{Generalization Initialization}
This section addresses the $I$-problem for initializing the generalization function. As aforementioned, $I$-problem can be regarded as a $K$-means clustering problem of users' private data with two generalization constraints. 
The original $K$-means clustering problem (without any constraint) is already NP-hard in a Euclidean space~\cite{Ckmeans1d}, so our $I$-problem is also NP-hard. To this end, we modify a widely-used heuristic $K$-means algorithm~\cite{Hartigan1979StatisticalAA} to satisfy both $(k,\alpha)$-uniqueness and $(l,\beta)$-variety given a cluster number $K$. Note that $K$ should be set to $\frac{|U|}{\alpha} \le K \le \frac{|U|}{k}$, where $|U|$ is the number of users.

The $K$-means algorithm~\cite{Hartigan1979StatisticalAA} is initiated by randomly selecting a center for each cluster; it then repeats the following two steps until a stopping criterion is achieved: 1) assign users' private data to their closest clusters according to the Euclidean distances between the data and the clusters' centers; 2) re-identify the center for each cluster obtained by step one. To further include the constraints of $(k,\alpha)$-uniqueness and $(l,\beta)$-variety, we design an adapted constraint-aware $K$-means approach. Basically, to assign data samples to clusters, the proposed constraint-aware algorithm prioritizes the low-bound requirements that each cluster at least includes $k$ users and $l$ exact private attribute values. In particular, instead of assigning a data sample to the closest cluster directly as the $K$-means method, our method adds a data sample to the closest cluster that has not yet possessed $k$ users or $l$ exact private attribute values. In other words, if a cluster includes enough data samples that satisfy the lower-bound requirements, it will not be assigned more samples until all the clusters satisfy the lower-bound requirements. Furthermore, for maintaining data utility, our method refuses assigning a data sample to a cluster if the cluster would exceed $\alpha$ users or $\beta$ exact private attribute values. Once a data sample is assigned to a cluster, we adjust the cluster center for the next round of data sample assignments.  Algorithm~\ref{alg:i-problem} depicts the detail initial generalization process.
\begin{algorithm}[t!]
	\footnotesize
	\SetKwInOut{Input}{Input}
	\SetKwInOut{Output}{Output}
	\Input{
		$Y$: private attribute values;\\
		$K$: number of generalization groups;\\
		$k$ and $\alpha$: $(k,\alpha)$-uniqueness constraint;\\
		$l$ and $\beta$: $(l,\beta)$-variety constraint;\\
	}
	\Output{$\mathcal G_{\widetilde Y|Y}^{0}$: initial generalization function;
	}
	$\widetilde Y \leftarrow  \varnothing$; $\widetilde Y_c \leftarrow \varnothing$; $Y^- \leftarrow Y$\;
	\While{$true$}{
	    \For{$g$ in $[1,K]$}{
	    \uIf{initial}{$y^c_g \leftarrow \text{Rand}(Y^-)$; //randomly select a center for group $y_g$}
	    \Else{$y^c_g \leftarrow \text{Center}(y_g)$; //re-identify the group center}
	    $y_g\leftarrow\{y^c_g\}$; //initialize group $y_g$\ with the center $y^c_g$\\
	    $Y^- \leftarrow Y^-\setminus \{y^c_g\}$; //the remaining ungrouped private values\\
	    $\widetilde Y \leftarrow \widetilde Y \cup \{y_g\}$; //initial generalized groups\\
	    $\widetilde Y_c \leftarrow \widetilde Y_c \cup \{y^c_g\}$; //the center set of initial generalized groups
	    }
	    $S_{k,l} \leftarrow  \widetilde Y$; $S_{\alpha,\beta} \leftarrow  \widetilde Y$\;
	    \While{$Y^-\neq \varnothing$}{ 
	    //$d_{min}(Y^-,\widetilde Y_c,S)$ searches a pair of an exact private value $y$ and a group $y_g$ from $S$ that attains the smallest distance and satisfies $|U_y|+|U_{y_g}|\le \alpha$\\
    	    \uIf{$S_{k,l}\neq \varnothing$}{
    	    $y, y_g \leftarrow d_{min}(Y^-,\widetilde Y_c,S_{k,l})$;
    	    }\uElseIf{$S_{\alpha,\beta}\neq \varnothing$}{
    	    $y, y_g \leftarrow d_{min}(Y^-,\widetilde Y_c,S_{\alpha,\beta})$;
	        }\Else{
	        set $K \leftarrow K+1$ and restart the algorithm\;
	        \Return\;}
    	    $y_g \leftarrow y_g \cup \{y\}$\;
    	    $Y^- \leftarrow Y^-\setminus \{y\}$\;
    	    \If{$|U_{y_g}|\ge k \text{ and }|y_g|\ge l$}{$S_{k,l}\leftarrow S_{k,l}\setminus \{y_g\}$;}
    	    \If{$|U_{y_g}| = \alpha$ or $|y_g| = \beta$}{$S_{\alpha,\beta} \leftarrow S_{\alpha,\beta}\setminus \{y_g\}$\;}
	  }
		\If{convergence}{
			break\;
		}
	}
	$\mathcal G_{\widetilde Y|Y}^{0} \leftarrow$ the initial generalization function that maps $Y$ to $\widetilde Y$\;
	\Return{$\mathcal G_{\widetilde Y|Y}^{0}$}\;
	\caption{Initialization ($I$-problem)}
	\label{alg:i-problem}
\end{algorithm}

\subsection{Generalization-aware Obfuscation}
This section addresses $O$-problem by proposing a generalization-aware obfuscation approach. The goal is to learn an optimal $\mathcal{O}_{\hat X|X}$ for privacy-insensitive data $X$ by minimizing privacy leakage conditioned on a given $\mathcal{G}_{\widetilde Y|Y}$. While the objective in $O$-problem is convex given $\widetilde Y$, it cannot be easily solved when the user size is larger than a few hundred. Therefore, instead of learning the obfuscation function based on individual users, we turn to optimize the obfuscation function on user clusters.

Specifically, we first cluster users based on their privacy-insensitive data. Each user is mapped into a user cluster and our individual user-based obfuscation function $\mathcal{O}_{\hat X|X, \widetilde Y}$ becomes a user cluster-based obfuscation function $\mathcal{O}_{\hat C|C, \widetilde Y}$. In principle, such a convex optimization problem with several constraints can be solved by many solvers (\eg, CVX~\cite{CVX_2008}) with a computational complexity of $O(|C|^2)$, where $|C|$ is the number of user clusters (see Algorithm~\ref{alg:o-problem}).
\begin{algorithm}[t!]
	\footnotesize
	\SetKwInOut{Input}{Input}
	\SetKwInOut{Output}{Output}
	\Input{
		$\mathcal G_{\widetilde Y|Y}$: generalization function;\\
		$\Delta Q_{c}$: utility loss budget;\\
	}
	\Output{$\mathcal O_{\hat C|C,\widetilde Y}$: obfuscation function;
	}
	$C \leftarrow$ cluster users based on their privacy-insensitive data;\\
	$\widetilde Y \leftarrow$ generalize Y according to $\mathcal G_{\widetilde Y|Y}$\;
	$\mathcal O_{\hat C|C,\widetilde Y} \leftarrow$: solve the following optimization problem:
	\begin{align*}
		\min \limits_{\mathcal{O}_{\hat C|C, \widetilde Y}} \ \ & I(\hat C, \widetilde Y)  \\
		\text{s.t.}\ \ \ \ \ \   &  \mathbb{E}_{\tilde y\in \widetilde Y} \mathbb{E}_{C \hat C}[d(C, \hat C| \tilde y)] \le \Delta Q_C  \\
		& \mathcal{O}_{\hat C|C}(\hat c|c) \in [0,1], \forall c,\hat c\in \mathcal{C}  \\
		& \sum \limits_{\hat c}  \mathcal{O}_{\hat C|C}(\hat c|c)=1,\forall c\in \mathcal{C}
	\end{align*}\\	
	\Return{$\mathcal O_{\hat C|C,\widetilde Y}$;}
	\caption{Obfuscation ($O$-problem)}
	\label{alg:o-problem}
\end{algorithm}

\subsection{Generalization with Stochastic Privacy-Utility Boosting}
To address the $G$-problem and minimize the privacy leakage in Eq.~\eqref{eq:GsubobjFun} given the optimized $\hat X$ of the $O$-problem, we propose a stochastic privacy-utility boosting algorithm to further lower the privacy leakage by searching for a better $\widetilde Y$. 

Because the objective function of the $G$-problem is not convex, we cannot solve it directly. Instead, we first adjust the prior generalization function $\mathcal{G}_{\widetilde Y'|Y}$ by modifying a stochastically selected generalized value.  A new candidate generalization function $\mathcal{G}_{\widetilde Y|Y}$ is generated if the generalized private data given $\mathcal{G}_{\widetilde Y|Y}$ still satisfy $(k,\alpha)$-uniqueness and $(l,\beta)$-variety. Then we compute privacy leakage and data utility loss given $\hat X$ and the new generalization function $\mathcal{G}_{\widetilde Y|Y}$. Ideally, while $\hat X$ is fixed and two privacy requirements are ensured, the new $\mathcal{G}_{\widetilde Y|Y}$ is better than the prior $\mathcal{G}_{\widetilde Y'|Y}$ if both privacy leakage and data utility loss are reduced. The better generalization function $\mathcal{G}_{\widetilde Y|Y}$ will be used for the next iteration of $O$-problem solving. 

How to search for a new candidate generalization function is critical for designing a stochastic algorithm. Given a prior generalized results $\widetilde Y'$, we define $\tilde{y}'_n \in \widetilde Y'$  as a neighbor of a generalized value $\tilde y' \in  \widetilde Y'$ if $\tilde y'$  and $\tilde{y}'_n$ are adjacent in terms of private values. For instance, age stages 10-20 and 21-25 are  neighbors as ages 20 and 21 are adjacent. On this basis, we can generate a new generalization function by moving the boundary between $\tilde y'$ and $\tilde{y}'_n$ with a stochastically selected direction (\eg, modify age stages to  10-19 and 20-25). 
In practice, we randomly select a generalized private value $\tilde y'$ and one of its neighbors $\tilde{y}'_n$ for generating new candidates (see Algorithm~\ref{alg:g-problem}).
\begin{algorithm}[t!]
	\footnotesize
	\SetKwInOut{Input}{Input}
	\SetKwInOut{Output}{Output}
	\Input{ $\mathcal G_{\widetilde Y'|Y}$: the prior generalization function;\\
		$\mathcal O_{\hat C|C,\widetilde Y}$: the optimal obfuscation function given $\mathcal G_{\widetilde Y'|Y}$;\\
		$k$ and $\alpha$: $(k,\alpha)$-uniqueness constraint;\\
		$l$ and $\beta$: $(l,\beta)$-variety constraint;\\
	}
	\Output{$\mathcal G_{\widetilde Y|Y}$: better generalization function for $\mathcal O_{\hat C|C,\widetilde Y}$;
	}
	$I^{0} \leftarrow$: calculate privacy leakage given $\mathcal O_{\hat C|C,\widetilde Y}$ and $\mathcal G_{\widetilde Y'|Y}$\;
	$Q^{0} \leftarrow$: calculate data utility loss given $\mathcal O_{\hat C|C,\widetilde Y}$ and $\mathcal G_{\widetilde Y'|Y}$\;
	
	\While{$true$}{
		$\mathcal G_{\widetilde Y|Y}^{c} \leftarrow$ find a new candidate generalization function that can meet  $(k,\alpha)$-uniqueness and $(l,\beta)$-variety constraints;\\
		$I^c \leftarrow$ calculate privacy leakage given $\mathcal O_{\hat C|C,\widetilde Y}$ and $\mathcal G_{\widetilde Y|Y}^{c}$\;
		$Q^c \leftarrow$ calculate data utility loss given $\mathcal O_{\hat C|C,\widetilde Y}$ and $\mathcal G_{\widetilde Y|Y}^{c}$\;
		\If{$I^c < I^0$ and $Q^c < Q^0$}{
			$Q^0 \leftarrow Q^c,\, I^0 \leftarrow I^c,\, G_{\widetilde Y|Y} \leftarrow G_{\widetilde Y|Y}^{c}$;
		}
		\If{$I^c$ and $Q^c$ convergence}{
			break \;
		}
	}
	
	\Return{$\mathcal G_{\widetilde Y|Y}$;}
	\caption{Generalization ($G$-problem)}
	\label{alg:g-problem}
\end{algorithm}

\subsection{Data Obscuring and Theoretic Properties}
\label{sec:DataObscuring}
Algorithm~\ref{alg:hyobscure} summarizes the procedure of \textit{HyObscure}, which addresses the hybrid privacy-preserving data obscuring problem by three steps: 1) initializing $\mathcal G^0_{\widetilde Y|Y}$ by solving $I$-problem; 2) solving $O$-problem and $G$-problem in a cross-iterative manner until convergence; 3) with the obtained $\mathcal G_{\widetilde Y|Y}$ and $\mathcal O_{\hat C|C}$, generating $\widetilde Y$ and $\hat X$ for publishing.

Specifically, in the last step, we first generalize $Y$ by $\mathcal G_{\widetilde Y|Y}$. Then, for any specific user $u$ with generalized private data $\tilde y$ and privacy-insensitive data $x_u$, we obtain $u$'s cluster $C$, and probabilistically sample an obfuscated cluster $\hat C$ by $\mathcal{O}_{\hat C|C, \widetilde Y}$. Finally we randomly select a user $u'$ with the same $\tilde y$ in $\hat C$ and obfuscate $x_u$ to $x_{u'}$.

\begin{algorithm}[t!]
	\footnotesize
	\SetKwInOut{Input}{Input}
	\SetKwInOut{Output}{Output}
	\Input{privacy-insensitive data $X$ and private data $Y$;}
	\Output{$\hat X$: obfuscated  $X$, $\widetilde Y$: generalized $Y$;}
	$\mathcal G_{\widetilde Y|Y} \leftarrow$ obtain  $\mathcal G_{\widetilde Y|Y}^0$ by solving $I$-problem\;
	\While{$not \  converged$}{
		$\mathcal O_{\hat C|C} \leftarrow$ solve $O$-problem given $\mathcal G_{\widetilde Y|Y}$ \;
		$\mathcal G_{\widetilde Y|Y} \leftarrow$ solve $G$-problem given $\mathcal O_{\hat C|C}$ \;
	}
	$\hat X$,$\widetilde Y \leftarrow$ obfuscate $X$ by $\mathcal O_{\hat C|C}$, generalize $Y$ by $\mathcal G_{\hat Y|Y}$\;
	\caption{\textit{HyObscure}}
	\label{alg:hyobscure}
\end{algorithm}

For the convergence of HyObscure and the optimization performance, we have the following theorems (see proofs in Supplemental Document):

\begin{theorem}
	HyObsure converges in a finite number of iterations.~\hfill$\blacksquare$
\end{theorem}

\begin{theorem}
	The privacy leakage bound of HyObsure is
	\begin{equation}
	I(\hat X; Y|\mathcal{O}^*_{\hat X|X})+c,
	\end{equation}
	where $\mathcal{O}^*_{\hat X|X}$ is the optimized function when only obfuscation is concerned, and $c$ is a constant.~\hfill$\blacksquare$
\end{theorem}


\section{Evaluation}
\label{sec:experiment}
In this section, we conduct extensive experiments to evaluate HyObscure. All experiments are run on a Windows 10 desktop server with Intel Core i7-8700 CPU@3.2GHz and 16GB RAM.

\subsection{Experimental Setup}
\subsubsection{Data Sets} Two real-world datasets are leveraged for the experiments as follows.

\textbf{MovieLens} \cite{MovieLens1M}: The \textit{MovieLens100K} data set contains 943 users with their demographic information and ratings on 1,682 movies. Each user has rated at least 20 movies, and the total number of ratings is more than 100,000. We take movie ratings as privacy-insensitive data and age as private data for the experiments.

\textbf{Foursquare} \cite{yang2019privacy}: We use a Foursquare data set from New York City with 3,669 users and 893,722 check-ins at 1,861 POIs. Note that the home locations are sparsely distributed in New York City. Hence, we divide the area of New York City into $1km\times 1km$ grids and all the location points in one grid are equally represented by the center of the grid. In total, we have 2,640 grids. We employ users' daily check-in activities as privacy-insensitive data and their home address as private information.

\begin{figure*}[t!]
    \centering
    \begin{subfigure}[b]{0.23\linewidth}
        \includegraphics[width=\textwidth]{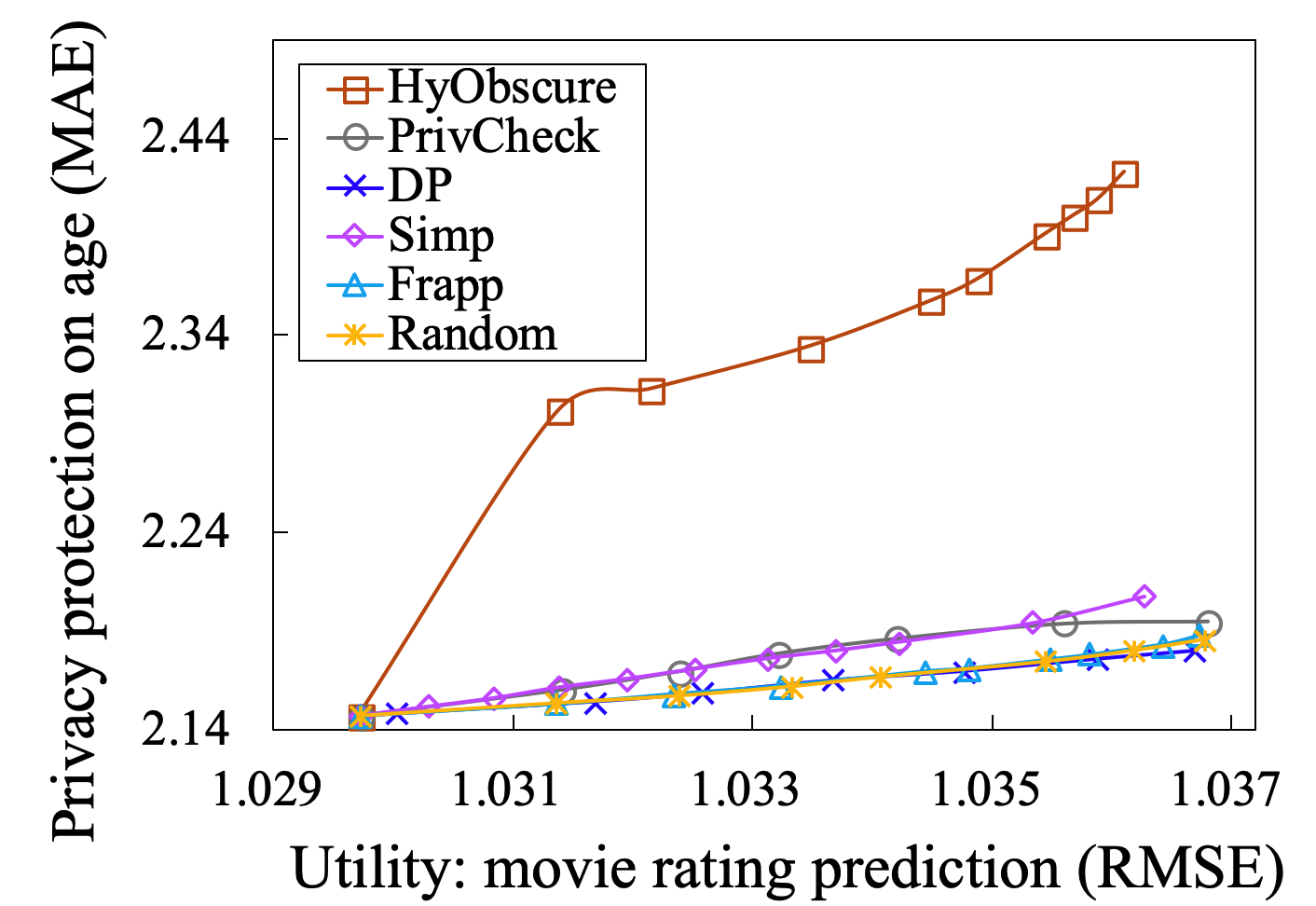}
        \caption{Scenario I by RF}
        \label{fig:AgeSIRF}
    \end{subfigure}
    \begin{subfigure}[b]{0.23\linewidth}
        \includegraphics[width=\textwidth]{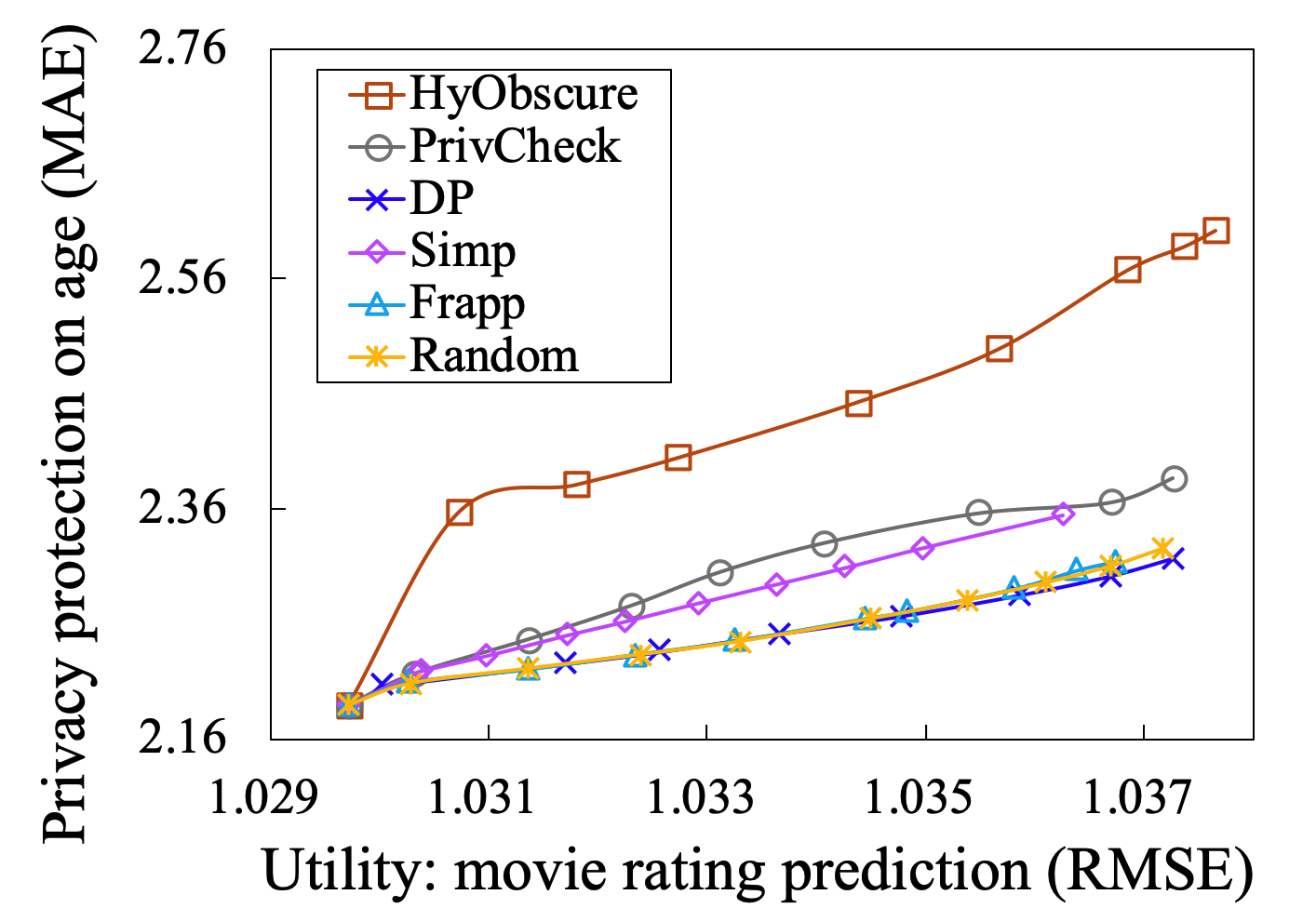}
        \caption{Scenario I by XGBoost}
        \label{fig:AgeSIXGBoost}
    \end{subfigure}
    \begin{subfigure}[b]{0.23\linewidth}
        \includegraphics[width=\textwidth]{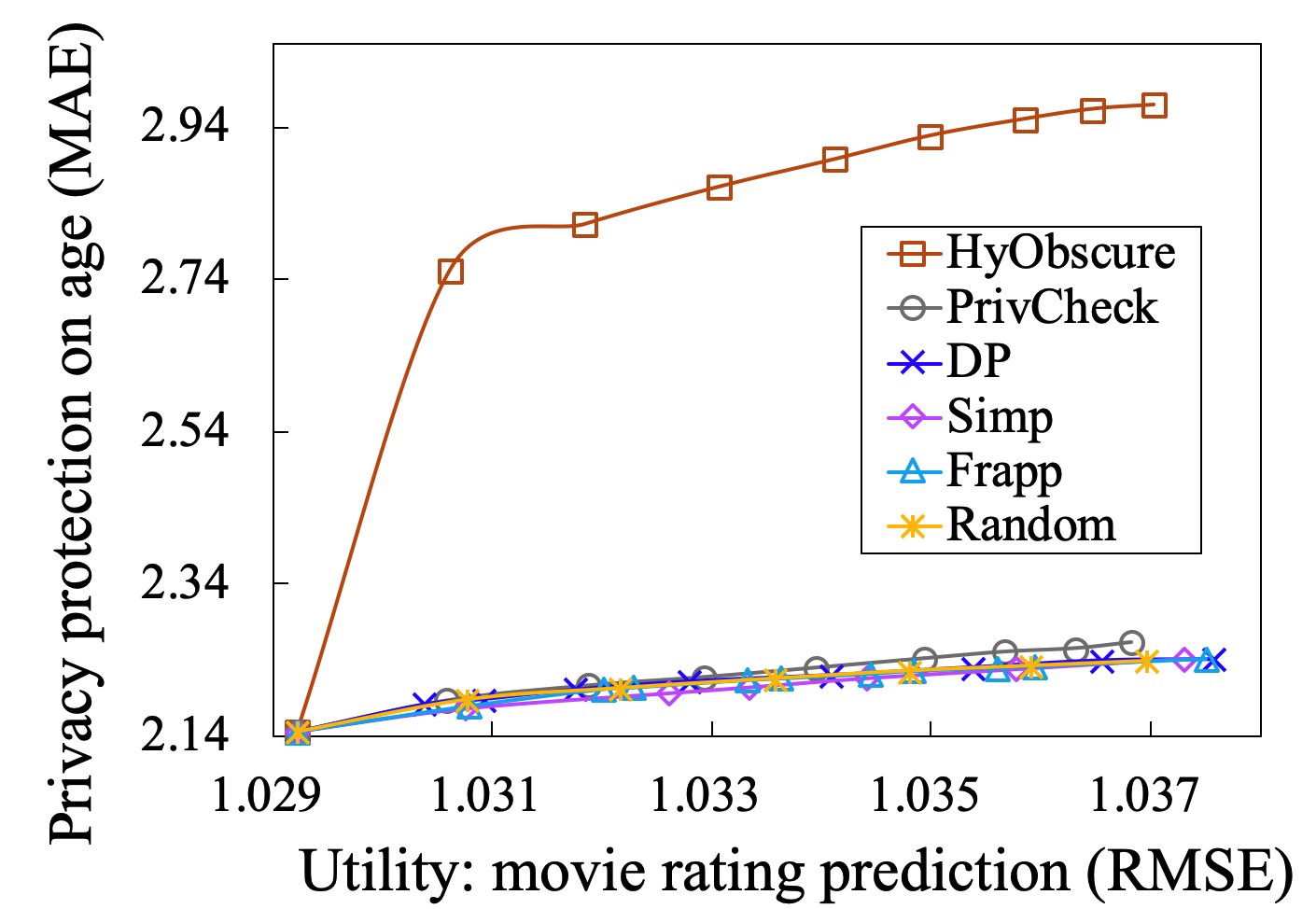}
        \caption{Scenario II by RF}
        \label{fig:AgeSIIRF}
    \end{subfigure}
     \begin{subfigure}[b]{0.23\linewidth}
        \includegraphics[width=\textwidth]{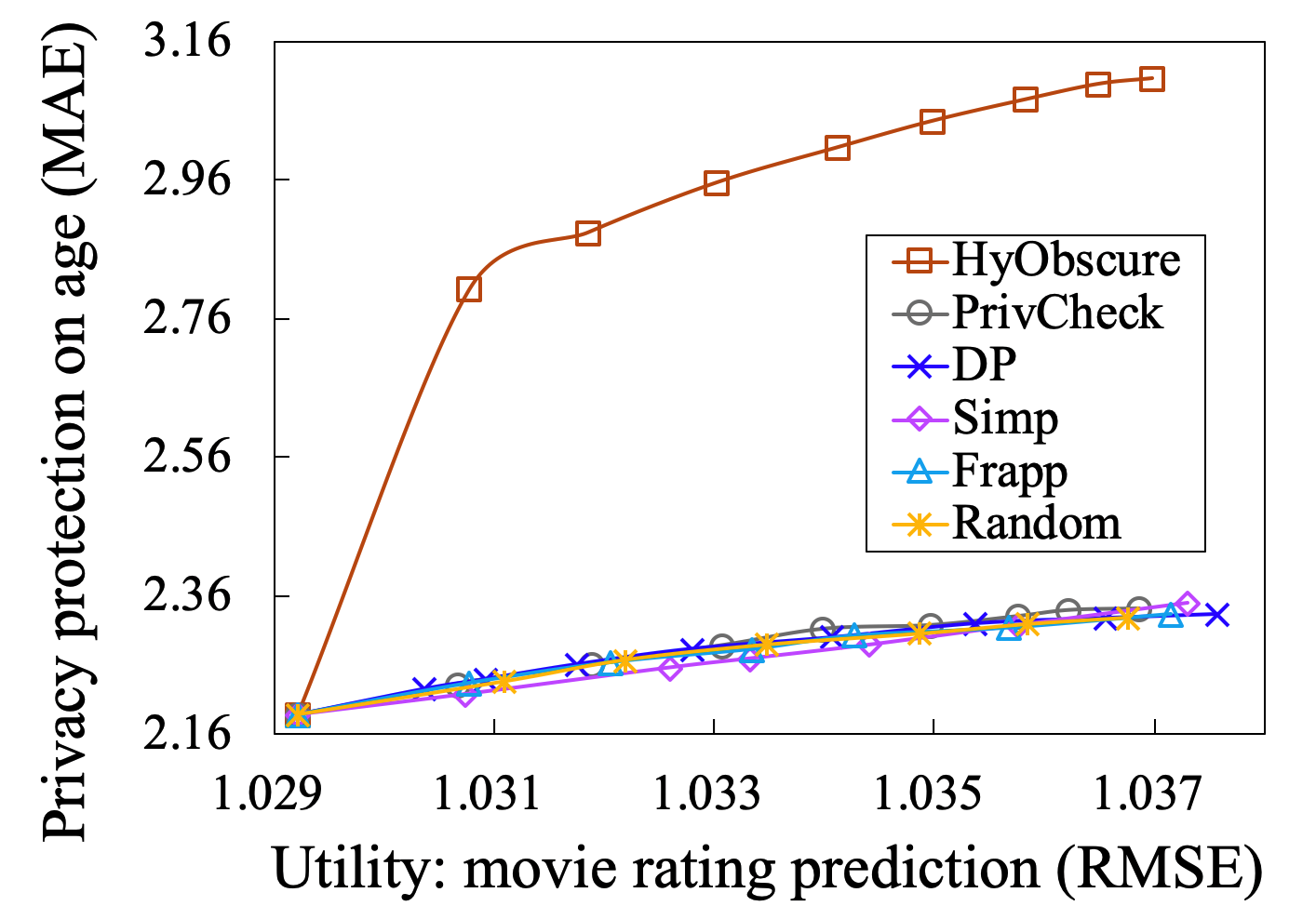}
        \caption{Scenario II by XGBoost}
        \label{fig:AgeSIIXGBoost}
    \end{subfigure}
    \caption{Tradeoff: privacy protection on \emph{age} vs. \emph{movie rating} prediction performance (MovieLens).}
    \label{fig:TradeOffage}
\end{figure*}

\begin{figure*}[t!]
    \centering
    \begin{subfigure}[b]{0.23\textwidth}
        \includegraphics[width=\textwidth]{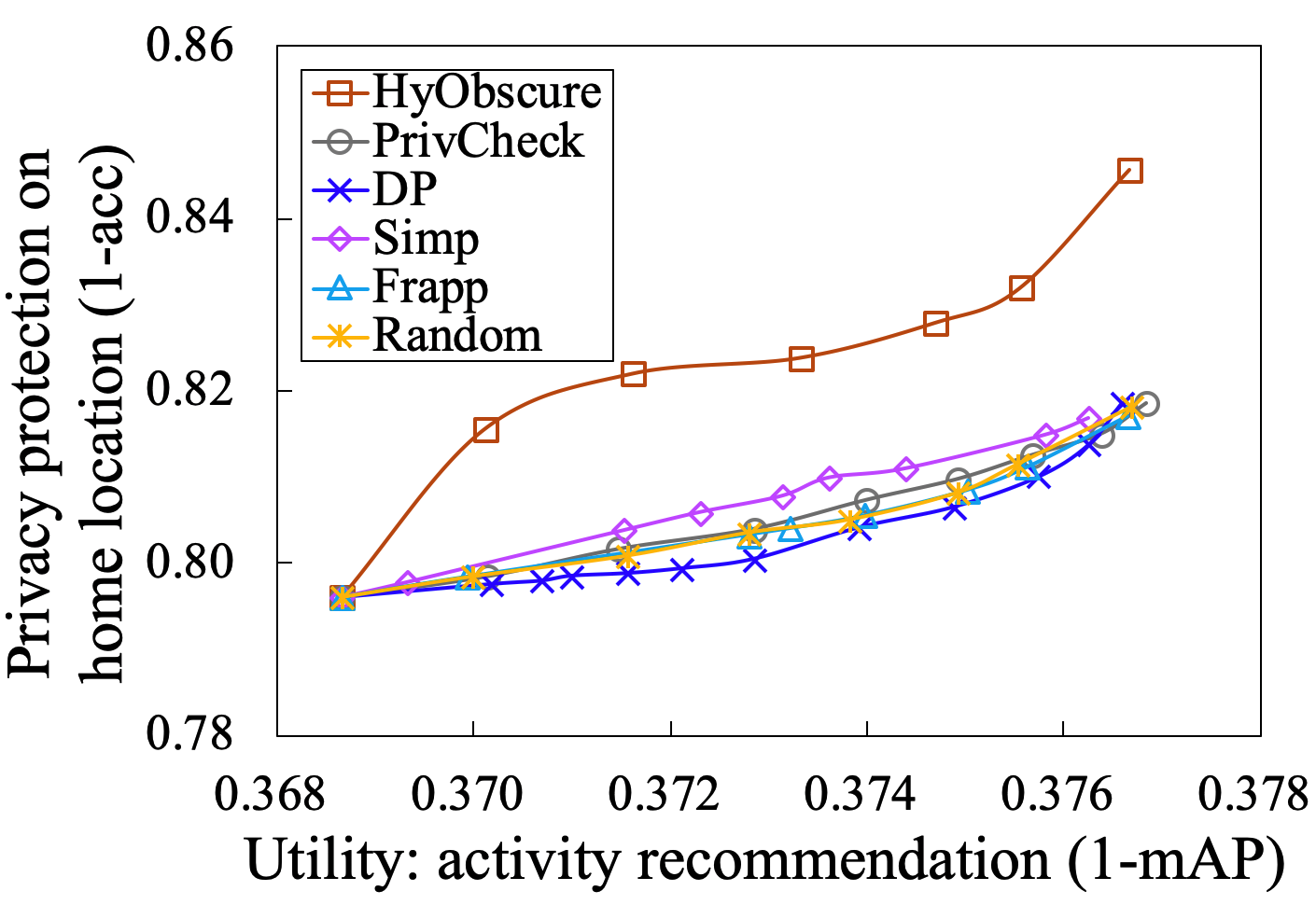}
        \caption{Scenario I by RF}
        \label{fig:HLSIRF}
    \end{subfigure}
    \begin{subfigure}[b]{0.232\textwidth}
        \includegraphics[width=\textwidth]{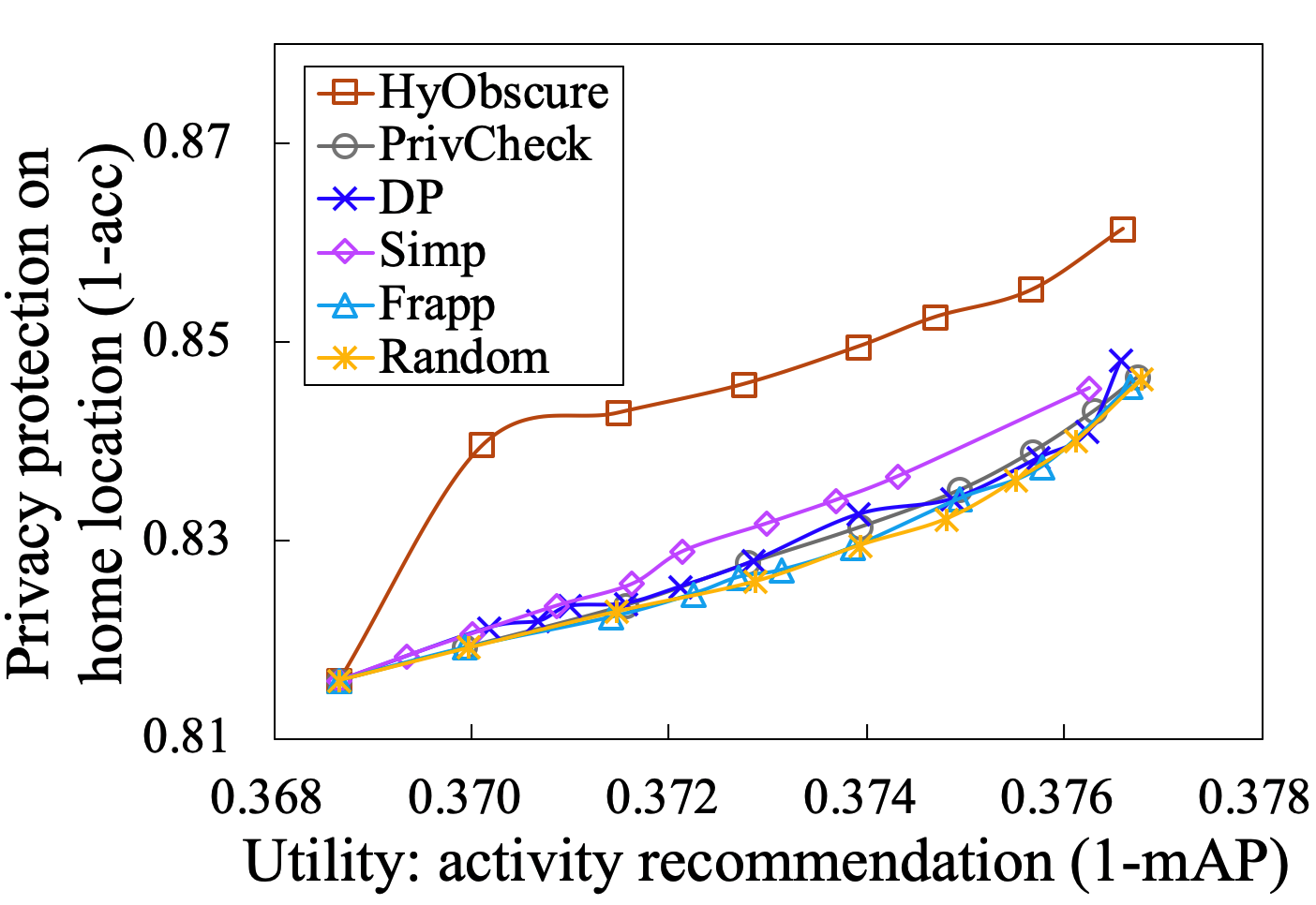}
        \caption{Scenario I by XGBoost}
        \label{fig:HLSIXGBoost}
    \end{subfigure}
    \begin{subfigure}[b]{0.232\textwidth}
        \includegraphics[width=\textwidth]{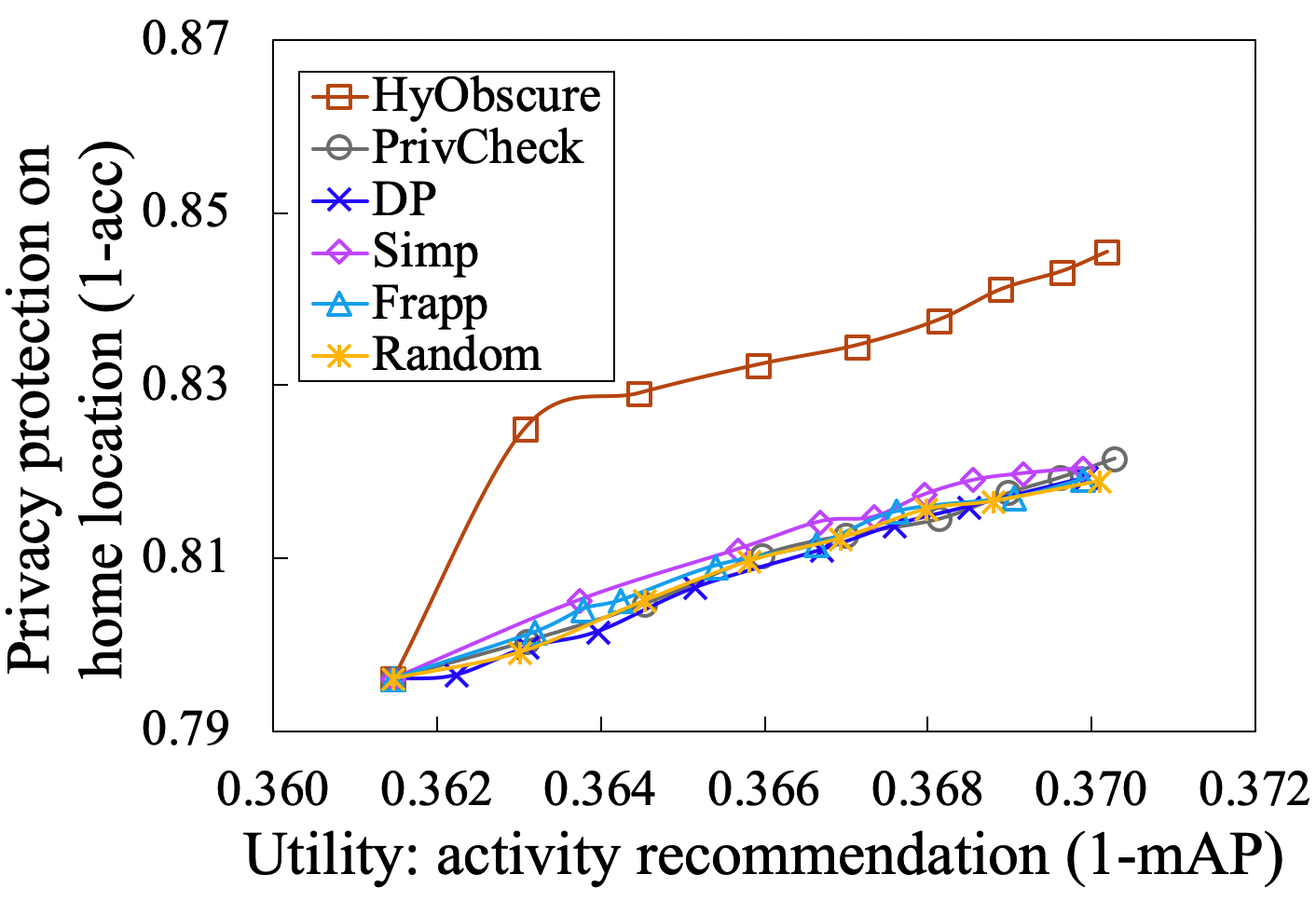}
        \caption{Scenario II by RF}
        \label{fig:HLSIIRF}
    \end{subfigure}
     \begin{subfigure}[b]{0.232\textwidth}
        \includegraphics[width=\textwidth]{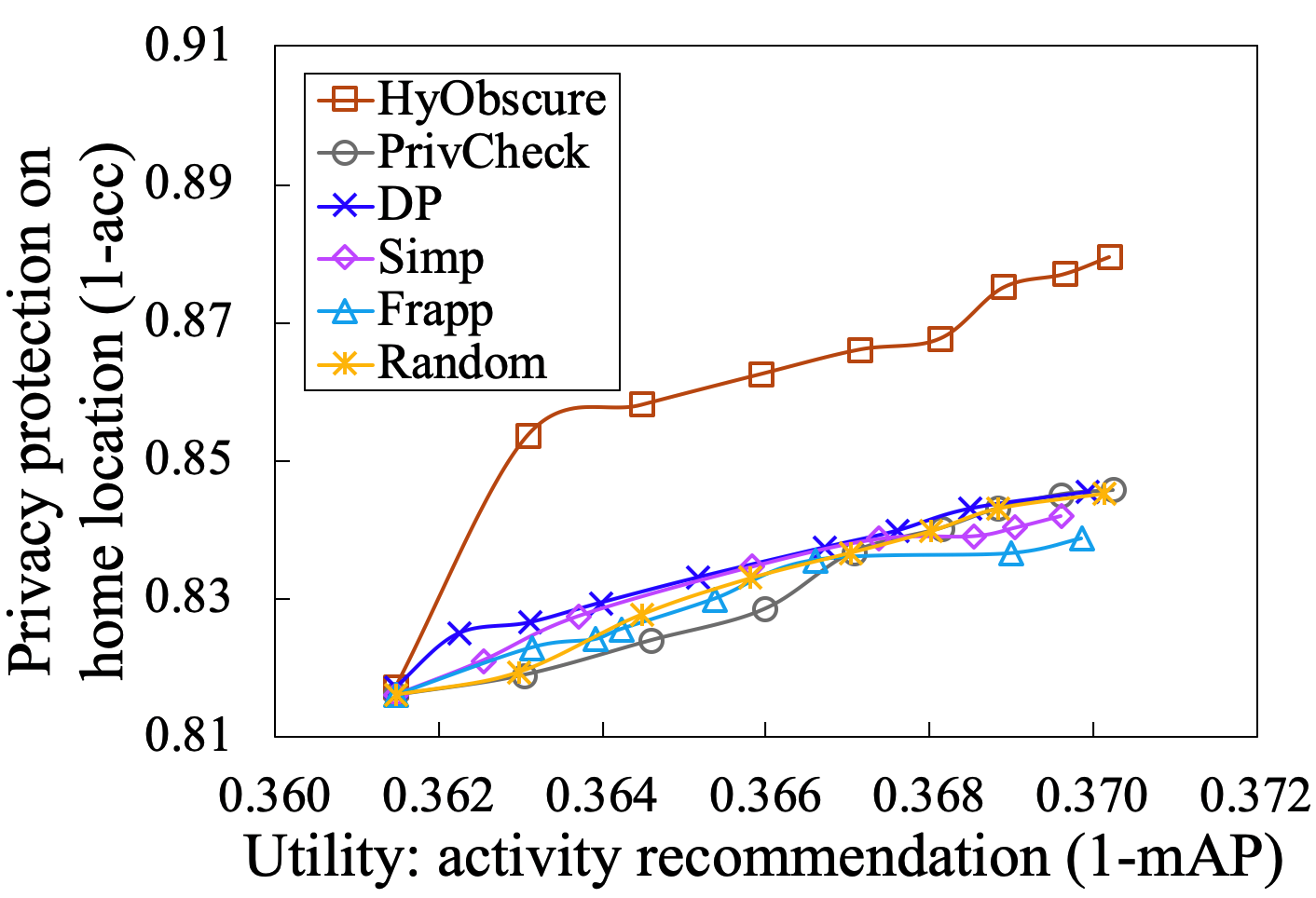}
        \caption{Scenario II by XGBoost}
        \label{fig:HLSIIXGBoost}
    \end{subfigure}
   \caption{Tradeoff: privacy protection on \emph{home address} vs. \emph{activity} recommendation performance (Foursquare).}\label{fig:TradeOfflocation}
   \vspace{-1em}
\end{figure*}

\subsubsection{Evaluation Scenarios}

We specify two real-life attribute inference attack scenarios used in our experiments:

\textbf{Scenario I}: We assume that the attackers derive a small number of users' real data including both privacy-insensitive and private information (\eg, \textit{real} movie ratings and \textit{exact} age), and train their attribute inference models based on the real data. The attribute inference models are then leveraged to predict other users' private data when their obscured data are publicly released.

\textbf{Scenario II}: We suppose that the attackers trade a set of obscured data with the data holders and meanwhile harvest a small number of users' real private data (\eg, \textit{obfuscated} movie ratings and \textit{exact} age). Note that the small number of users with real private data are a subset of users with obscured data. Then the attribute inference model can be learned on the small number of users whose obscured data and private data are both available to the attackers. It can be used to infer private data about users whose obscured data are accessible.

\subsubsection{Attribute Inference Attacks and Privacy Leakage}

The goal of attackers is to infer exact values about users' private data with their prior knowledge. 
For the sake of generality and efficiency, 
we adopt two representative and effective methods, i.e., Random Forest (RF)~\cite{Breiman01ML} and XGBoost~\cite{Chen16KDD}, to train the inference models.

\textbf{Age inference}. Age is regarded as a private attribute in MovieLens data set. We exploit users' movie ratings, either original (Scenario I) or obfuscated (Scenario II), and age stage (generalized age value) as inputs to train age inference attack models. We the age inference as a regression problem and measure the \textit{age} inference performance by \textit{mean absolute error} (MAE). Larger MAE indicates less privacy leakage and better privacy protection.

\textbf{Home address inference}.  With the Foursquare data set, we take users' check-in activities and areas of users' home addresses (generalized private value) as inputs to learn inference attack model regarding users' exact home addresses. We take home location inference as a multi-class classification problem and employ accuracy to assess its inference performance. A lower inference accuracy represents less privacy leakage. In this vein, we use $1$-\textit{Accuracy} ($1$-\textit{acc}) to assess the privacy protection ability on users' home addresses (larger values mean better protection).

\subsubsection{Data Utility in Applications} 
We exploit the published (obscured) data to carry out movie rating prediction and activity recommendation for data utility evaluation. 

\textbf{Movie rating prediction}. With the MovieLens data set, we rely on the obfuscated movie ratings and age stage to perform user-based collaborative filtering for movie rating prediction~\cite{konstan1997grouplens,herlocker1999algorithmic}. 
We use \textit{root mean square error} (RMSE) to assess the prediction performance. Lower RMSE indicates better prediction performance and higher data utility given the same prediction method.

\textbf{Activity recommendation}. 
Activity recommendation is implemented based on the analysis of users' personal check-in patterns and activity preference~\cite{Activity_Rec}. \textit{Mean average precision} (mAP) 
is often employed to estimate the recommendation performance. 
To make the value changing trend the same as RMSE for movie rating prediction, we use $1$-mAP to measure activity recommendation performance and data utility. In general, data with larger utility will lead to lower $1$-mAP and better recommendation performance.

\subsubsection{Baselines}
\begin{figure*}[htb]
    \centering
    \begin{subfigure}[b]{0.24\textwidth}
        \includegraphics[width=\textwidth]{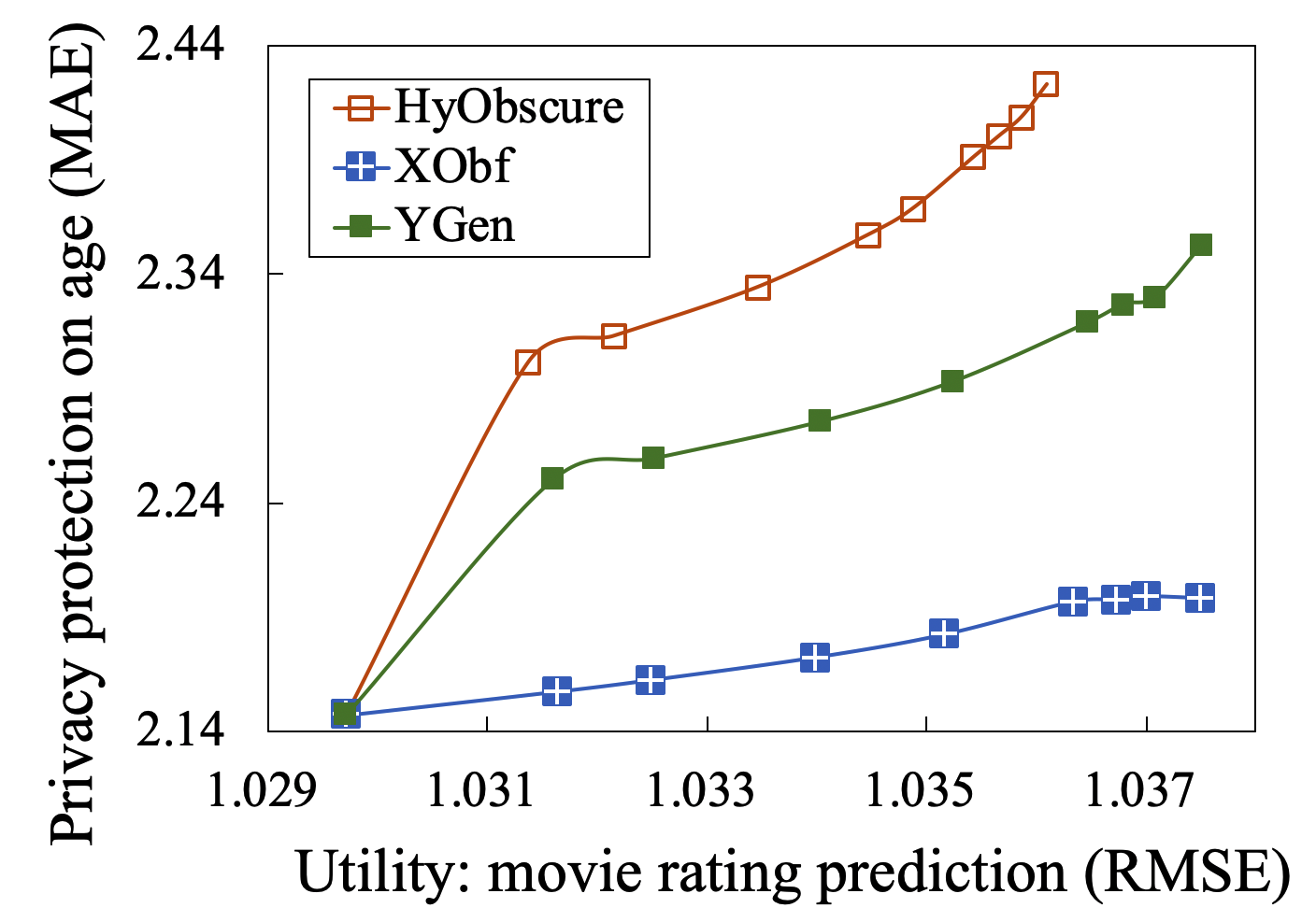}
        \caption{\emph{Age} in Scenario I}
        \label{fig:AgeSIRFComponent}
    \end{subfigure}\ 
    \begin{subfigure}[b]{0.24\textwidth}
        \includegraphics[width=\textwidth]{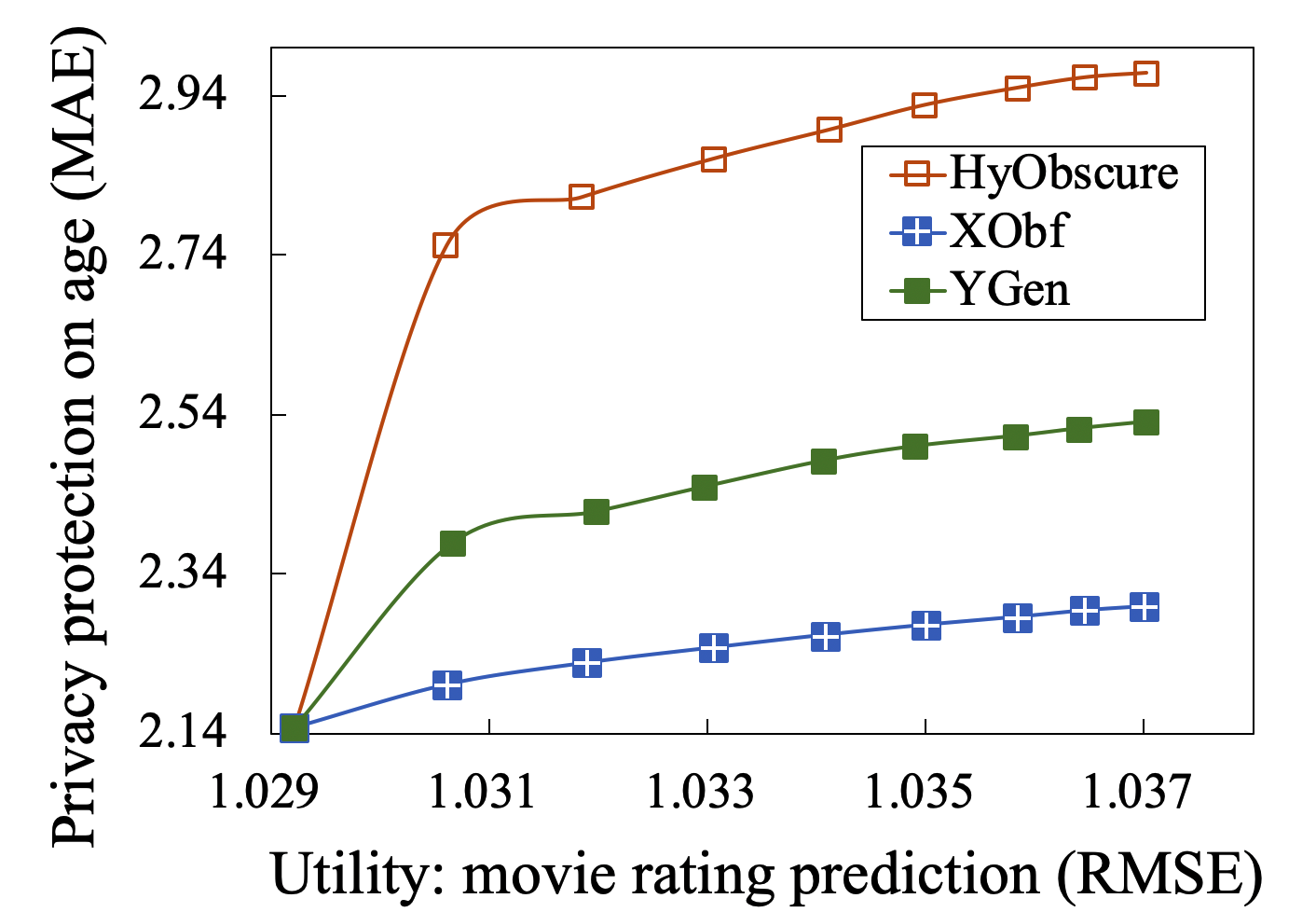}
        \caption{\emph{Age} in Scenario II}
        \label{fig:AgeSIIRFComponent}
    \end{subfigure}\ 
    \begin{subfigure}[b]{0.24\textwidth}
        \includegraphics[width=\textwidth]{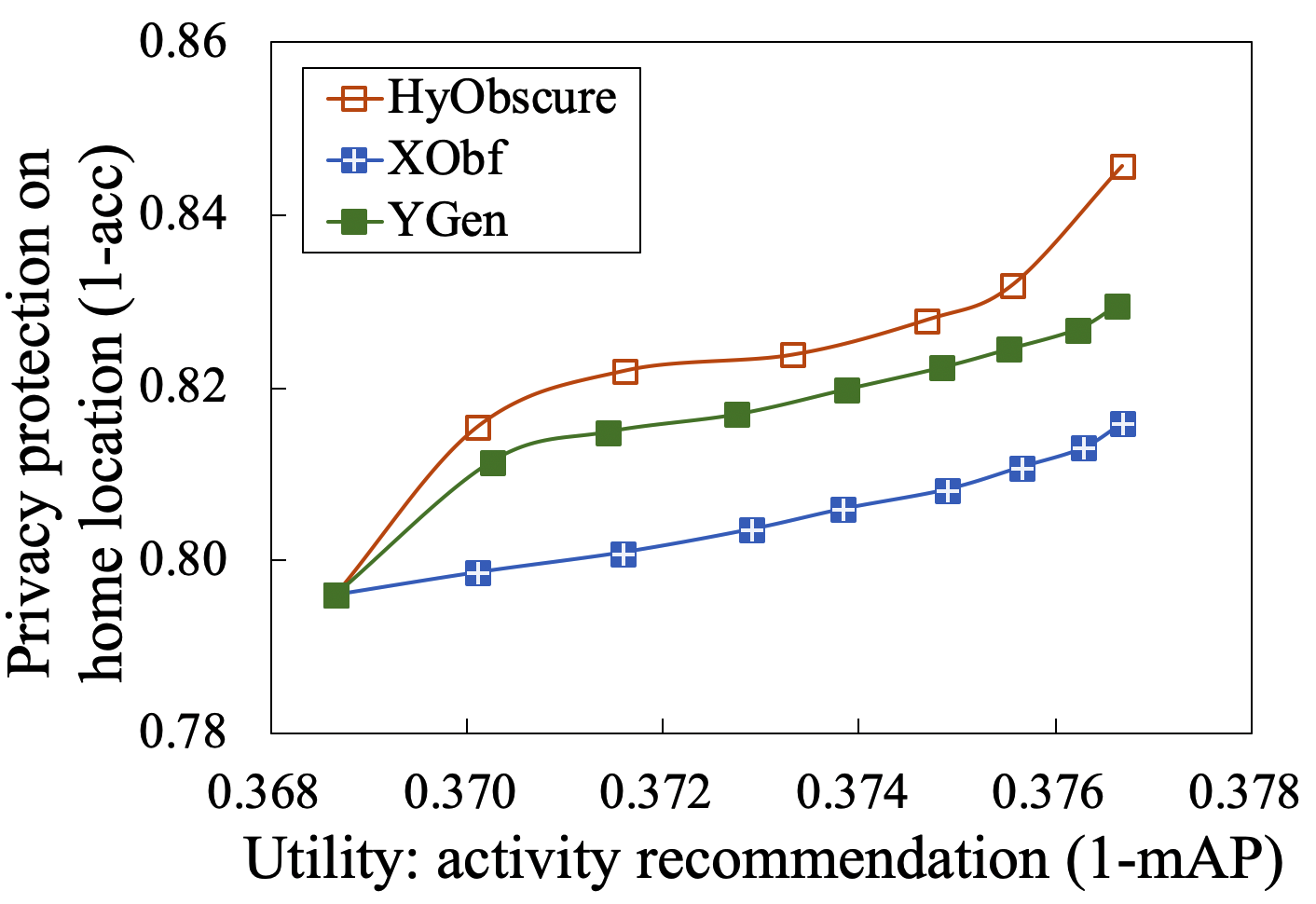}
        \caption{\emph{Location} in Scenario I}
        \label{fig:HLSIRFComponent}
    \end{subfigure}\ 
     \begin{subfigure}[b]{0.24\textwidth}
        \includegraphics[width=\textwidth]{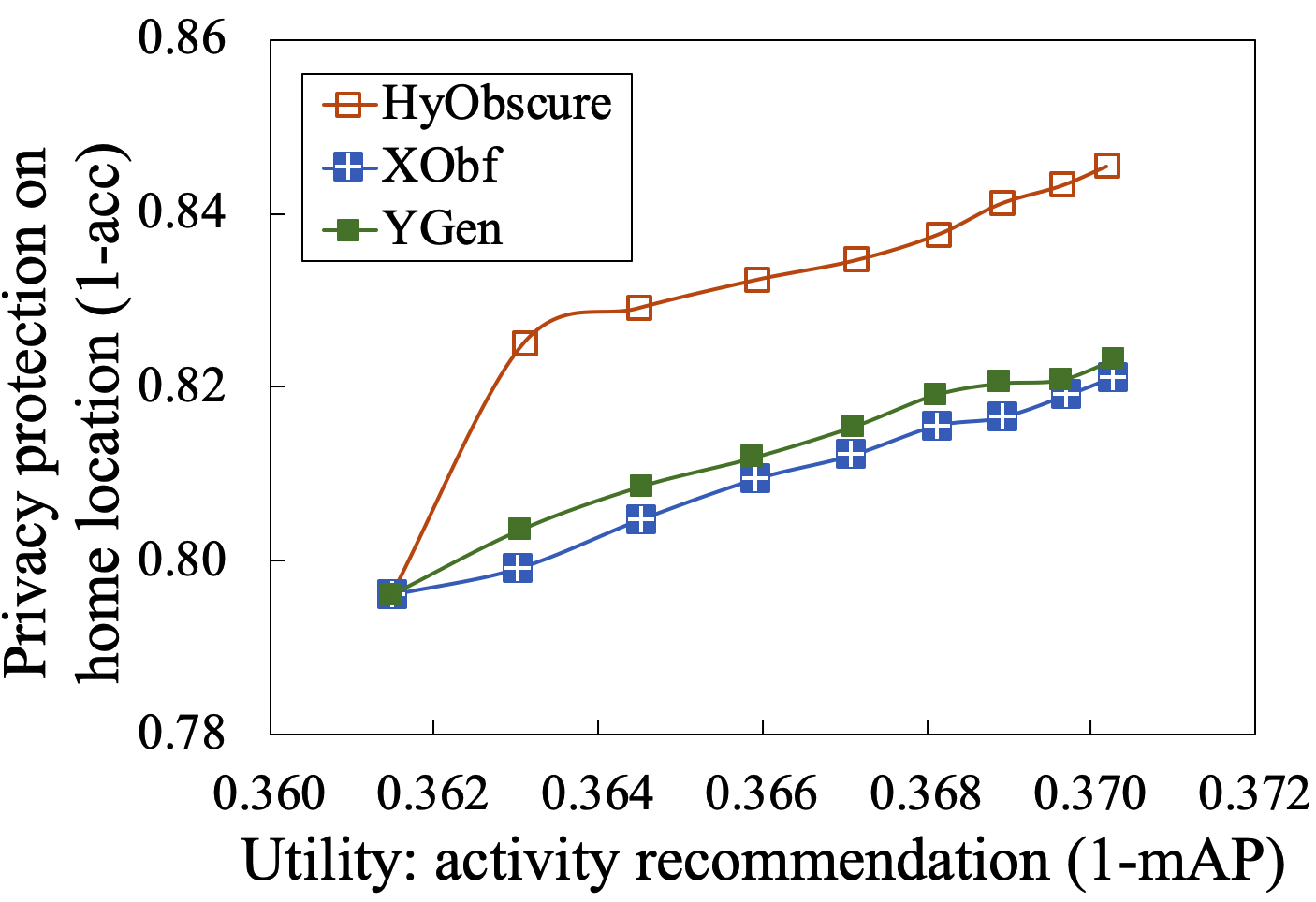}
        \caption{\emph{Location} in Scenario II}
        \label{fig:HLSIIRFComponent}
    \end{subfigure}
    \caption{Importance of cross-iterative optimization on obfuscation and generalization (RF).}\label{fig:CrossIterativeRF}
\end{figure*}

\begin{figure*}
    \centering
    \begin{subfigure}[b]{0.24\textwidth}
        \includegraphics[width=\textwidth]{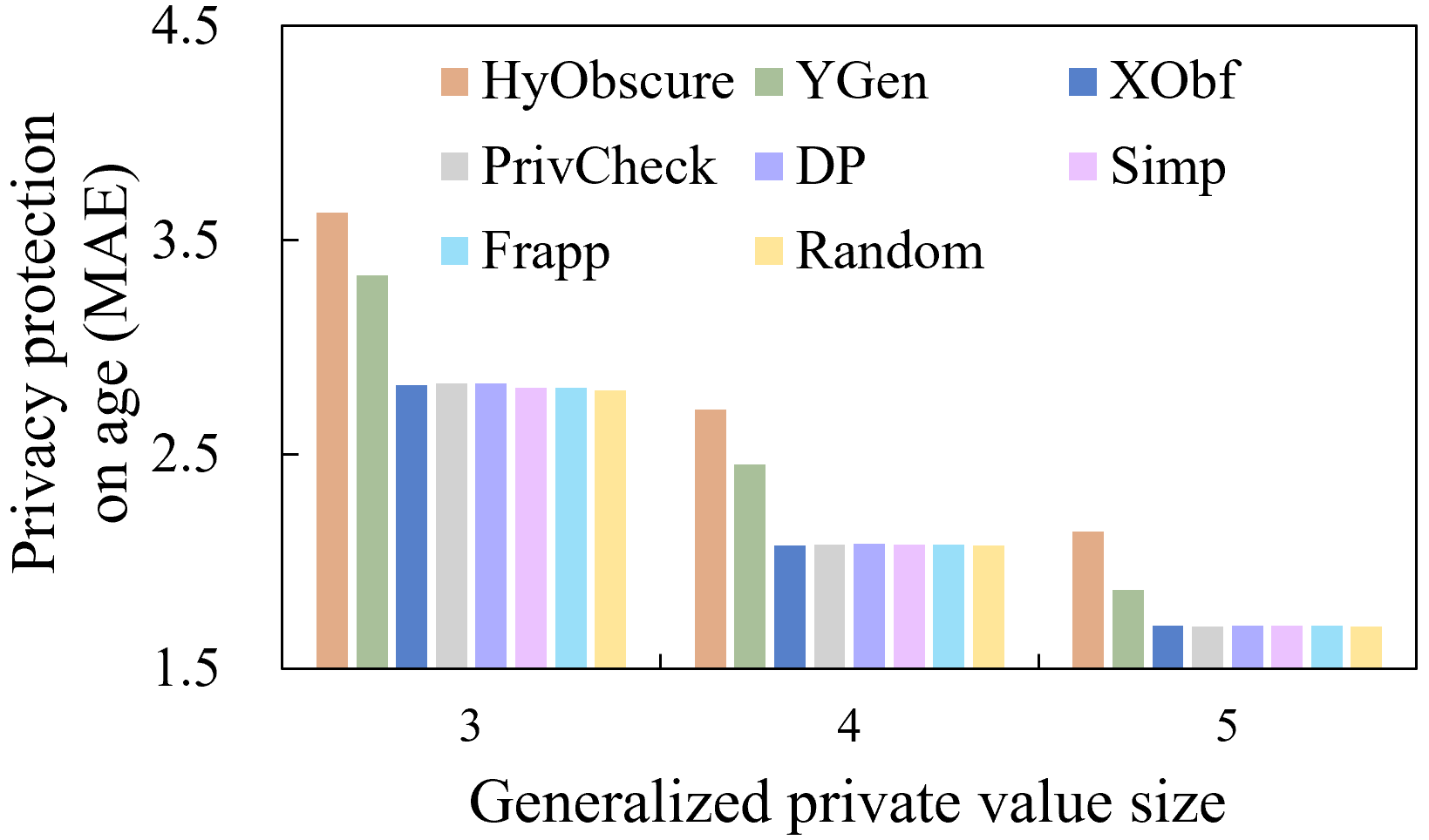}
        \caption{\emph{Age} in Scenario I}
        \label{fig:AgeSIRFGN}
    \end{subfigure}\ 
    \begin{subfigure}[b]{0.24\textwidth}
        \includegraphics[width=\textwidth]{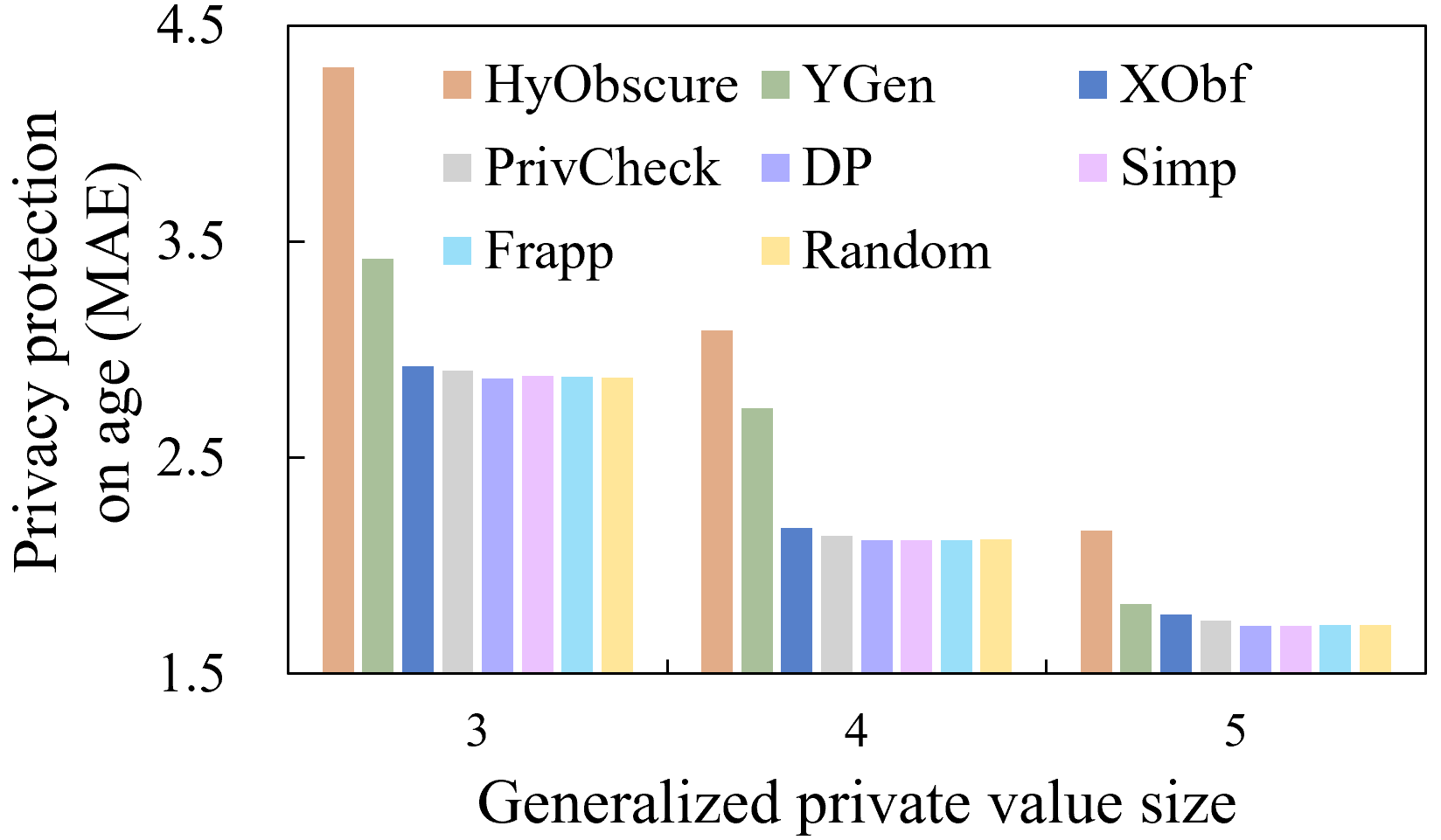}
        \caption{\emph{Age} in Scenario II}
        \label{fig:AgeSIIRFGN}
    \end{subfigure}\ 
    \begin{subfigure}[b]{0.24\textwidth}
        \includegraphics[width=\textwidth]{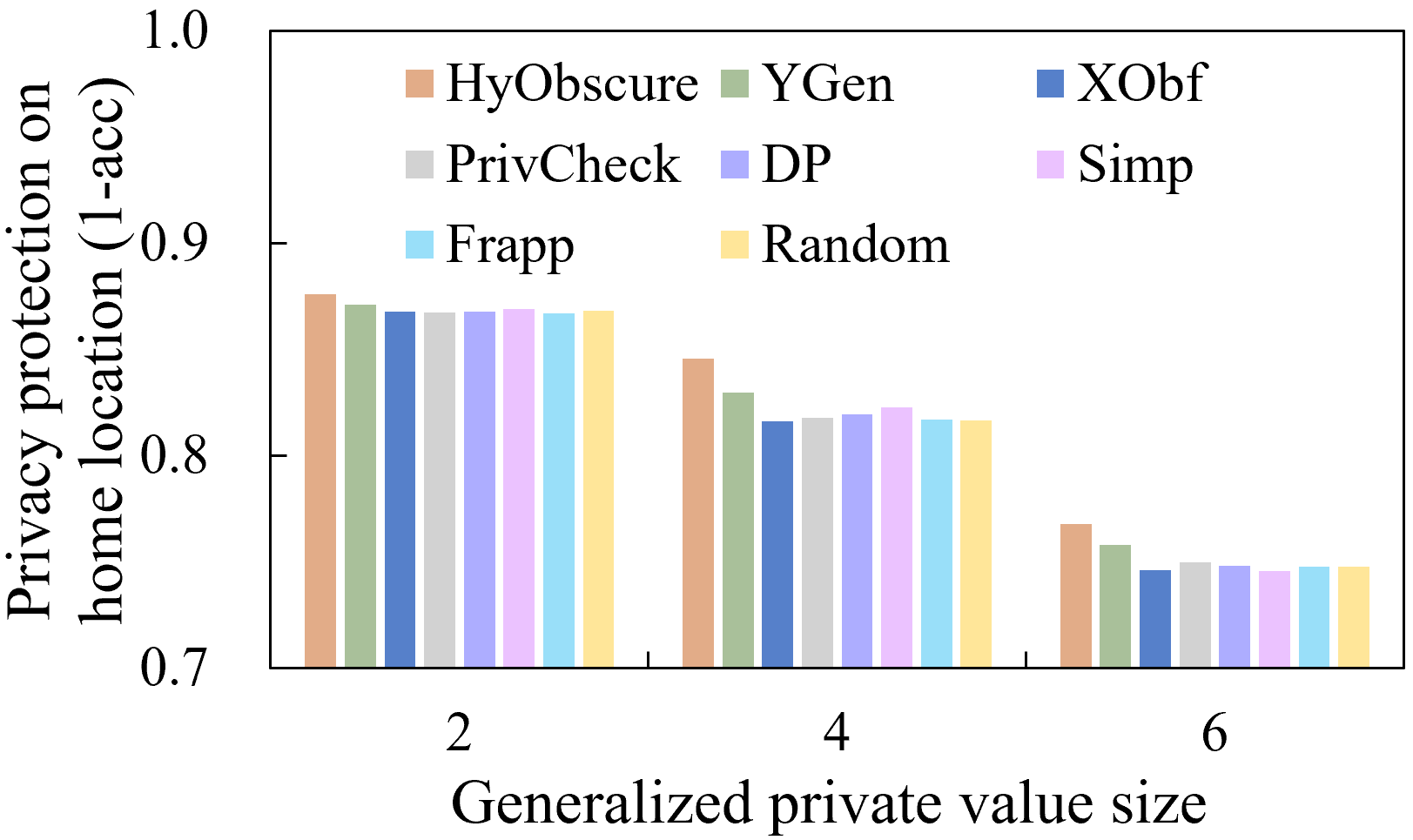}
        \caption{\emph{Location} in Scenario I}
        \label{fig:HLSIRFGN}
    \end{subfigure}\ 
    \begin{subfigure}[b]{0.24\textwidth}
        \includegraphics[width=\textwidth]{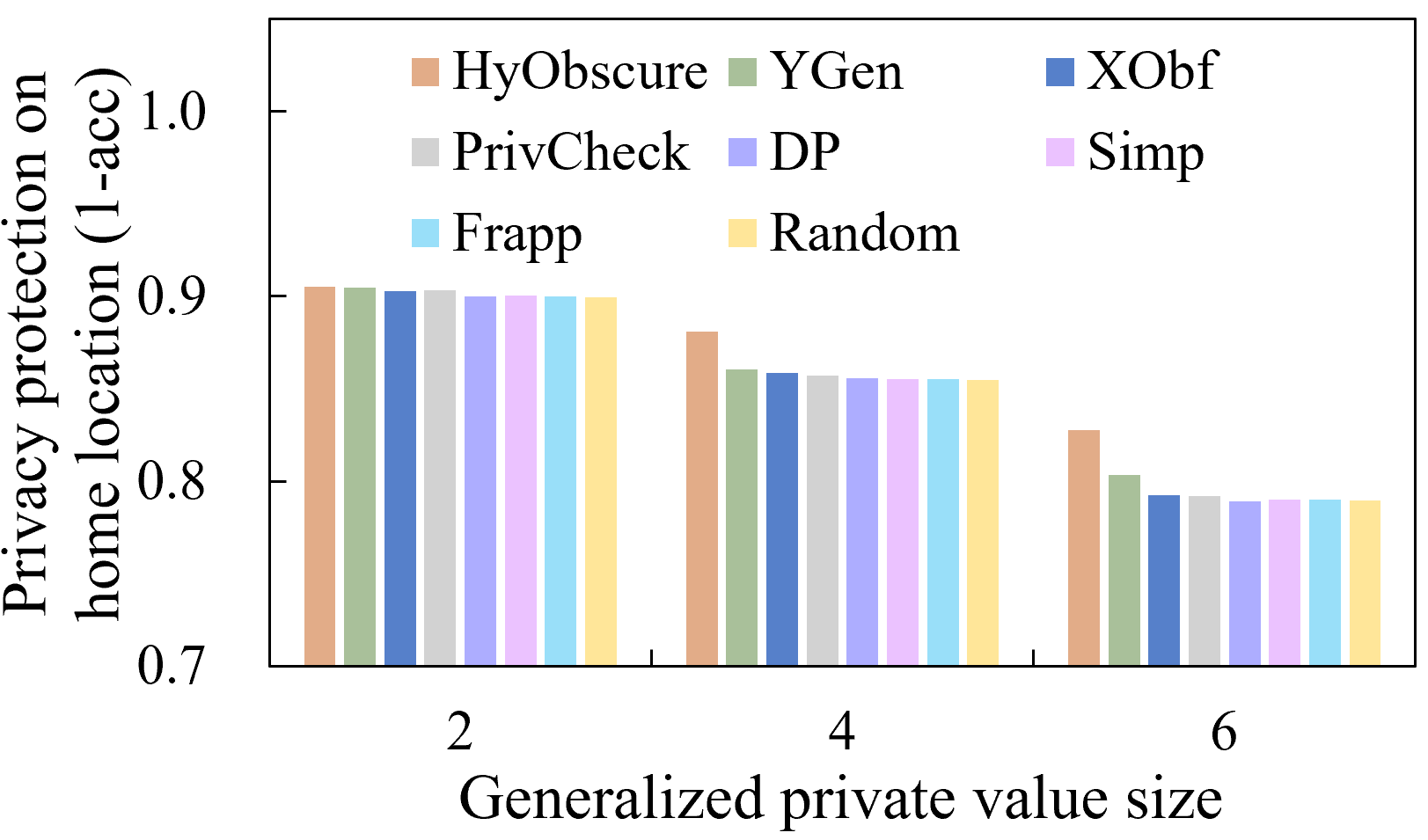}
        \caption{\emph{Location} in Scenario II}
        \label{fig:HLSIIRFGN}
    \end{subfigure}
    \caption{Impacts of generalized private value size on privacy protection given a certain utility (RF).}\label{fig:GNRF}
    \vspace{-1em}
\end{figure*}

Existing privacy-preserving data publishing approaches rarely consider hybrid obscuring for both privacy-insensitive and private data. To implement the baselines, we use the optimized initial generalization function for private data and perform state-of-the-art obfuscation methods for privacy-insensitive data without considering the data correlation. Specifically, we employ the following methods to obfuscate privacy-insensitive data as baselines and compare them with HyObscure.

$\bullet$ \textit{Random}~\cite{Yang2016PrivCheckPC}. It randomly selects some users for obfuscation given a default obfuscation budget and then obfuscates the users with another randomly selected user.


$\bullet$ \textit{Frapp}~\cite{agrawal2005framework}. It performs data obfuscation with a higher ratio to users themselves but with a less likelihood to others.

$\bullet$ \textit{Simp}~\cite{Yang2016PrivCheckPC}. 
It obfuscates data between users who are more similar (dissimilar) to each other with a larger (smaller) probability. 

$\bullet$ \textit{DP}~\cite{Wang2018GeographicDP}. It obfuscates user $u$ to $v$ by a probability decreasing exponentially with their distance $d(u,v)$, \ie, $p_{_{DP}}(v|u)\propto \exp(-\beta d(u,v))$. It satisfies $2\beta d_{max}$-differential privacy, where $d_{max} = \max \limits_{u,v\in U} d(u,v)$.

$\bullet$ \textit{PrivCheck}~\cite{Yang2016PrivCheckPC}. It optimizes an obfuscation function for privacy-insensitive data to achieve minimal privacy leakage given a distortion budget. 



In addition, we implement two variants of HyObscure to verify the importance of cross-iterative obscuring: 

$\bullet$ \textit{XObf}. Given the initial generalization approach $\mathcal{G}^0_{\widetilde Y|Y}$, \textit{XObf} carries out the generalization-aware obfuscation and aims to identify an optimal obfuscation function.

$\bullet$ \textit{YGen}. Given a obfuscation function $\mathcal{O}_{\hat X|X}$, \textit{YGen} runs the stochastic privacy-utility boosting approach to find a better generalization solution. \textit{PrivCheck} is set to the default $\mathcal{O}_{\hat X|X}$ as it performs the best among the baselines.

\begin{table}[t!]
	\centering
	\scriptsize
	\renewcommand*{\arraystretch}{1.1}
	\caption{Default setting of key parameters.}
	\vspace{-0.5em}
	\begin{tabular}{cccl}
		\toprule
		Notation       & MovieLens & Foursquare & \multicolumn{1}{c}{Description} \\ \midrule
		$(k,\alpha)$              & (100,200) & (100,150)           &      $(k,\alpha)$-uniqueness       \\
		$(l,\beta)$              & (4,8)   & (5,15)           &  $(l,\beta)$-variety    \\
		$|\widetilde Y| $& 4   &    4          &     Size of generalized private value  \\
		$|C|$        & 10  &    8          &  Size of user clusters\\
		\bottomrule
	\end{tabular}
	\label{tbl:defaultpar}
	\vspace{-1.5em}
\end{table}

\subsection{Experimental Results}
We report the experimental results of HyObscure from the following aspects: 1)~privacy-utility tradeoff, 2) effectiveness of cross-iterative obscuring, 3) impact of key parameters, 4) scalability, and 5) convergence. The default parameters set for the first two experiments are listed in Table~\ref{tbl:defaultpar}. Since all the experimental results are similar for the  RF and XGBoost inference attack models, we report the privacy-utility tradeoff using both RF and XGBoost models, and only report the results of RF for the rest experiments due to the page limit.

\subsubsection{Privacy-Utility Tradeoff}
We here conduct experiments using two different inference attack methods (RF and XGBoost) under two attack scenarios where attackers have different prior knowledge. Fig.~\ref{fig:TradeOffage} and Fig.~\ref{fig:TradeOfflocation} show the trend of protection effectiveness on private data along the varying retained data utility on MovieLens and Foursquare data sets. Note that 1) the increase of the number on $x$-axis (RMSE or 1-mAP) represents the decrease of utility, 2) a larger number on $y$-axis (MAE or 1-acc) indicates attackers make more errors on predicting private data and thus better privacy protection, and 3) the bottom-left most point in each figure is the reference point, indicating the ``worst'' privacy protection and the ``best'' utility with the unobscured original data.

Fig.~\ref{fig:TradeOffage} and  Fig.~\ref{fig:TradeOfflocation} generally show the same trend that the privacy protection performance is continuously improved with the decreasing of utility, demonstrating that the privacy protection indeed can be augmented by carefully obscuring data at the expense of utility reduction. More importantly, HyObscure achieves more efficient privacy-utility tradeoff by using lower utility loss to pursue better privacy protection effect. For instance, in Fig.~\ref{fig:AgeSIRF}, when the customized utility on movie rating prediction performance decreases around $0.2\%$ compared to the reference point, HyObscure can raise the privacy protection performance on \emph{age} by $7.2\%$, whereas the privacy protection increments by other baselines are only around $1\%$. Fig.~\ref{fig:HLSIRF} shows that when the privacy protection performance is around 0.817 in terms of 1-acc, the data utility loss caused by HyObscure is lower than the baseline approaches by more than $35\%$. Recall that all the baselines do not consider the correlation between privacy-insensitive and generalized private data when applying data obscuring, the results verify the importance of considering data correlation in hybrid data obscuring.

\subsubsection{Importance of Cross-iterative Optimization}
Fig.~\ref{fig:CrossIterativeRF} compares HyObscure with two variants on the privacy protection effect of the \emph{age} and \emph{location} data given certain levels of application utility. Note that both XObf and YGen subsequently optimize two obscuring functions without the cross-iterative process. It is obvious from Fig.~\ref{fig:CrossIterativeRF} that HyObscure delivers much higher privacy protection than that of XObf and YGen given the same level of utility. 
The results demonstrate that cross-iteratively optimizing the obfuscation function on privacy-insensitive data and the generalization function on private data can effectively strengthen the privacy protection capability of HyObscure.

\subsubsection{Impact of Key Parameters} 
Three groups of parameters are examined to verify the reliability of HyObscure. 

\textbf{Generalized private value size ($|\widetilde{\boldsymbol{Y}}|$)}.~We vary the generalized private value size to conduct a robustness check on HyObscure. In order to focus on the impact of $|\widetilde Y|$, we set default utility requirements and inspect whether HyObscure can perform better in terms of privacy protection at different levels of $|\widetilde Y|$. Fig.~\ref{fig:GNRF} displays the results where the RMSE of movie rating prediction is set to 1.038 for \emph{MovieLens} and the 1-mAP of activity recommendation is set to 0.371 for \emph{Foursquare}. The other parameters are set to default values as listed in Table~\ref{tbl:defaultpar}. These results verify that HyObscure consistently outperforms the baseline methods with varied $|\widetilde Y|$. Moreover, the results demonstrate that privacy protection becomes harder with the increase of $|\widetilde Y|$, \ie, more information is to be released. 

\begin{figure*}[t!]
	\centering
	\begin{subfigure}[b]{0.24\textwidth}
		\includegraphics[width=\textwidth]{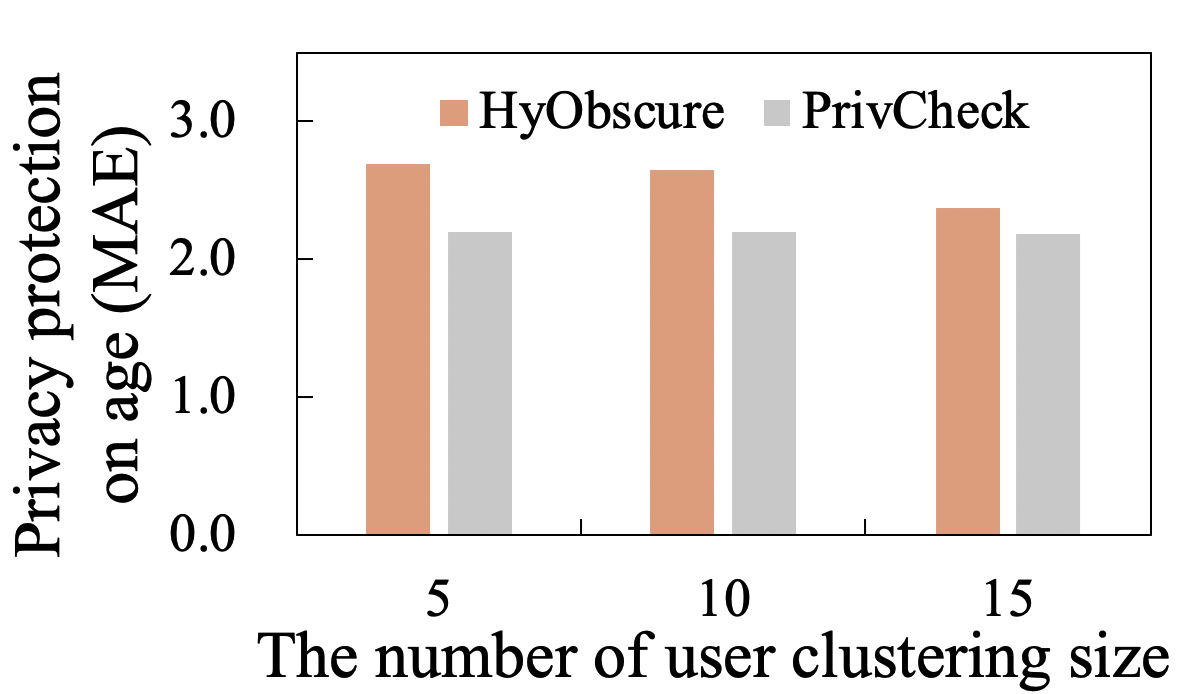}
		\caption{\emph{Age} in Scenario I}
		\label{fig:AgeSIRFCN}
	\end{subfigure}\ 
	\begin{subfigure}[b]{0.24\textwidth}
		\includegraphics[width=\textwidth]{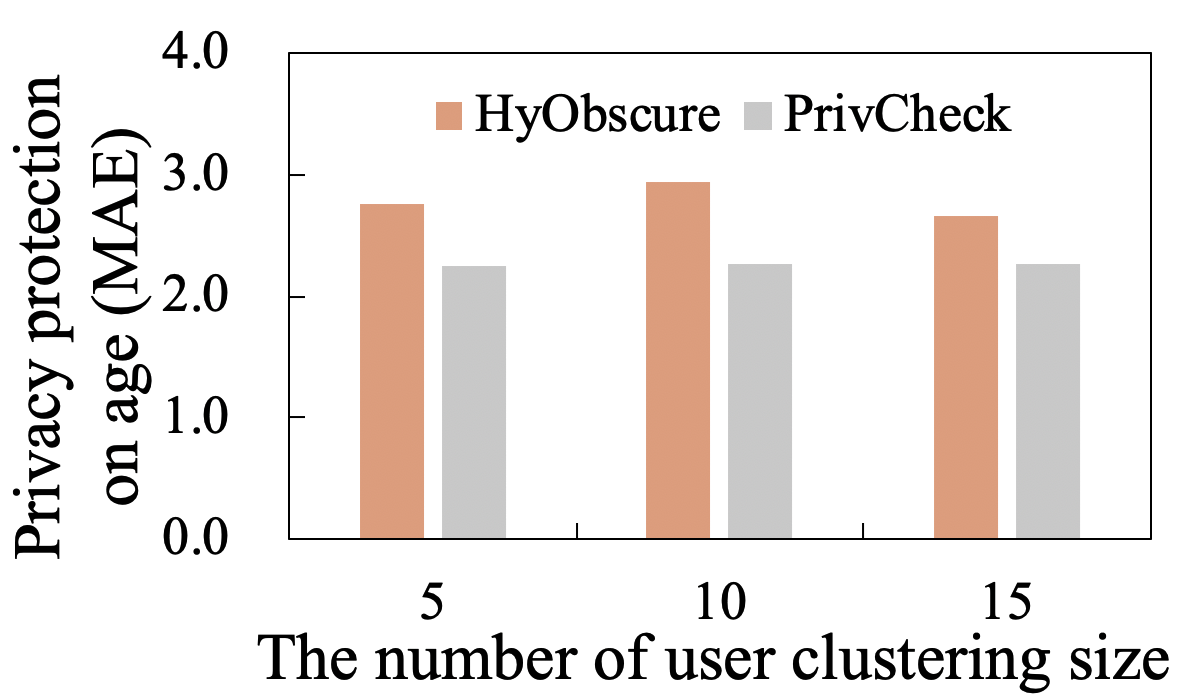}
		\caption{\emph{Age} in Scenario II}
		\label{fig:AgeSIIRFCN}
	\end{subfigure}\ 
	\begin{subfigure}[b]{0.24\textwidth}
		\includegraphics[width=\textwidth]{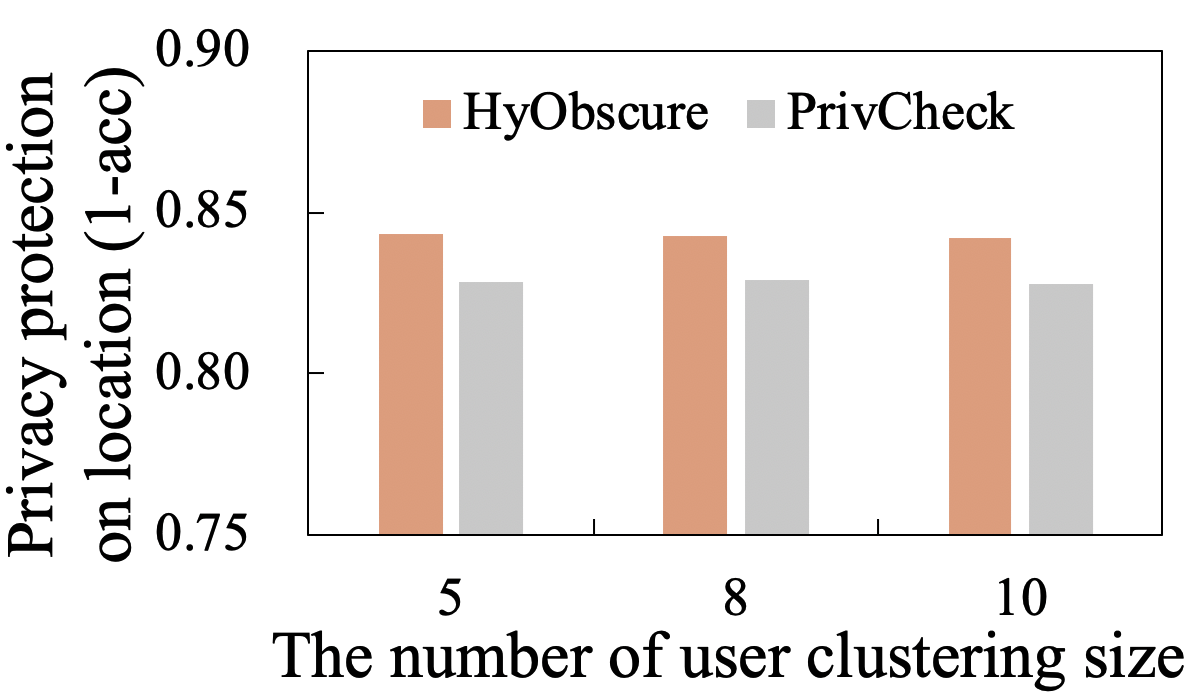}
		\caption{\emph{Location} in Scenario I}
		\label{fig:HLSIRFCN}
	\end{subfigure}\ 
	\begin{subfigure}[b]{0.24\textwidth}
		\includegraphics[width=\textwidth]{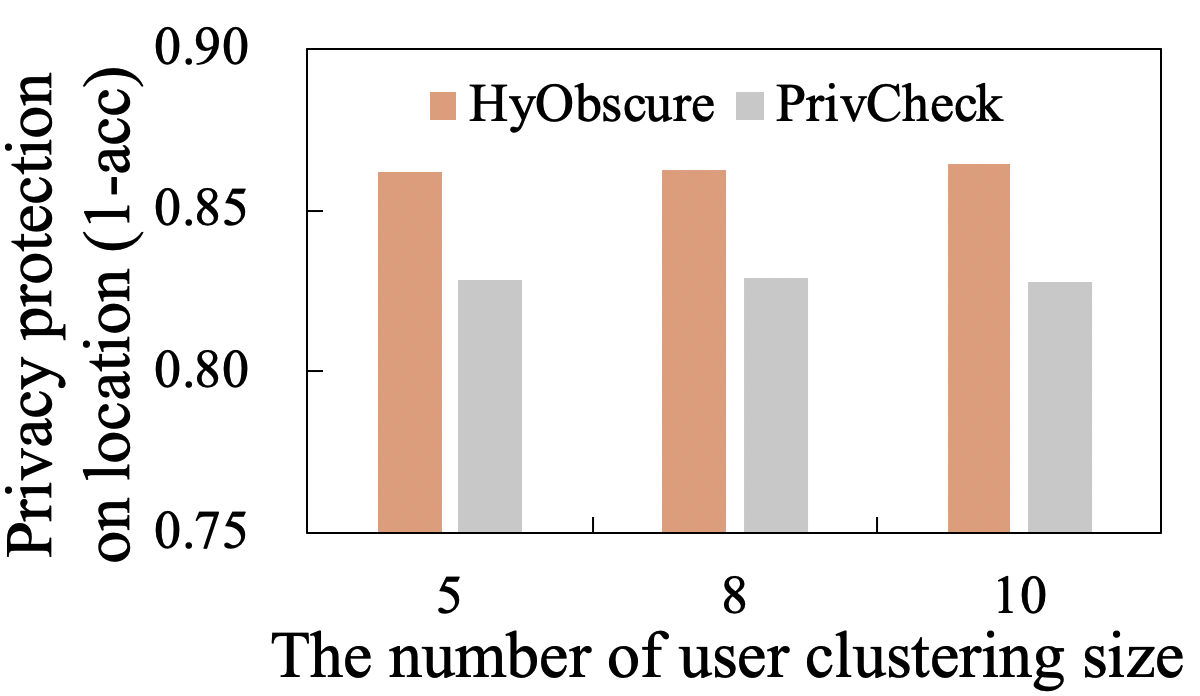}
		\caption{\emph{Location} in Scenario II}
		\label{fig:HLSIIRFCN}
	\end{subfigure}
	\caption{Impacts of the number of user clusters on privacy protection given a certain utility (RF).}\label{fig:CN}
	\vspace{-0.8em}
\end{figure*}

\begin{figure}[t!]
	\centering
	\begin{subfigure}[b]{0.47\linewidth}
		\includegraphics[width=\textwidth]{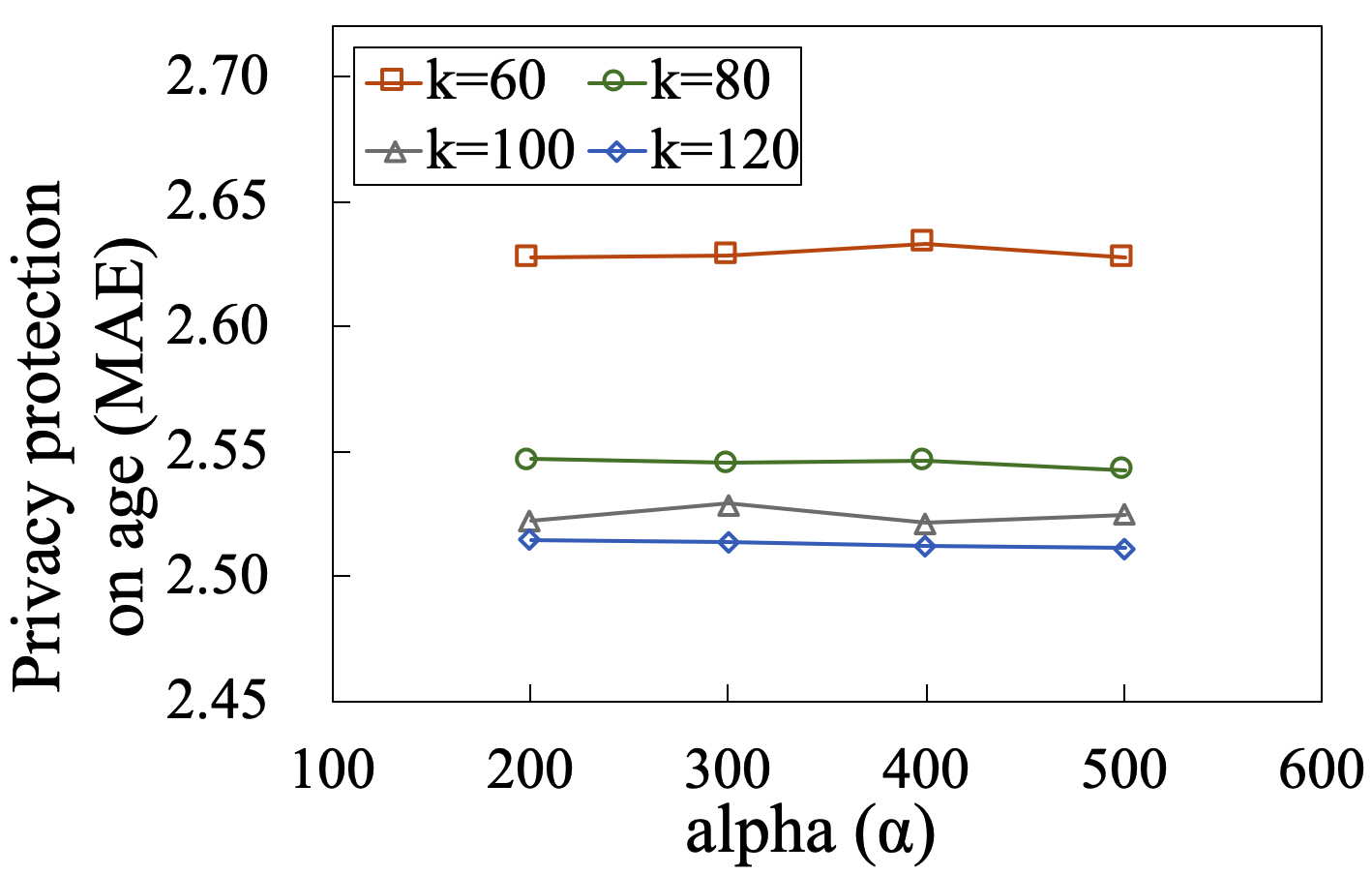}
		\caption{\emph{Age} (MovieLens)}
		\label{fig:ageKAlpha}
	\end{subfigure}\ \ 
	\begin{subfigure}[b]{0.47\linewidth}
		\includegraphics[width=\textwidth]{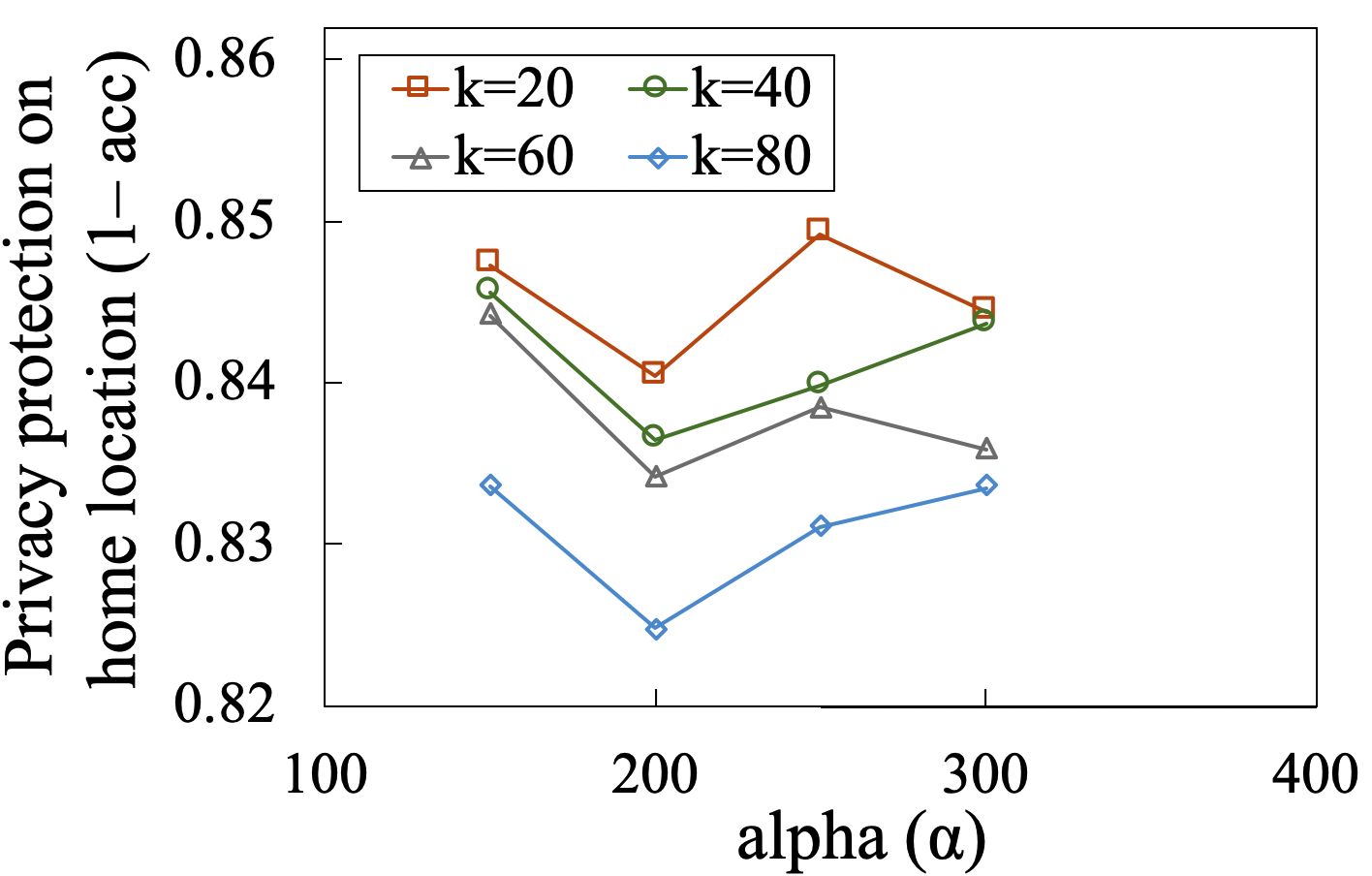}
		\caption{\emph{Location} (Foursquare)}
		\label{fig:locationKAlpha}
	\end{subfigure}
	\caption{Protection by $(\boldsymbol{k},\boldsymbol{\alpha})$-uniqueness (Scenario~I, RF).}
	\label{fig:kalpha}
	\vspace{-0.5em}
\end{figure}

\begin{figure}[t!]
	\centering
	\begin{subfigure}[b]{0.47\linewidth}
		\includegraphics[width=\textwidth]{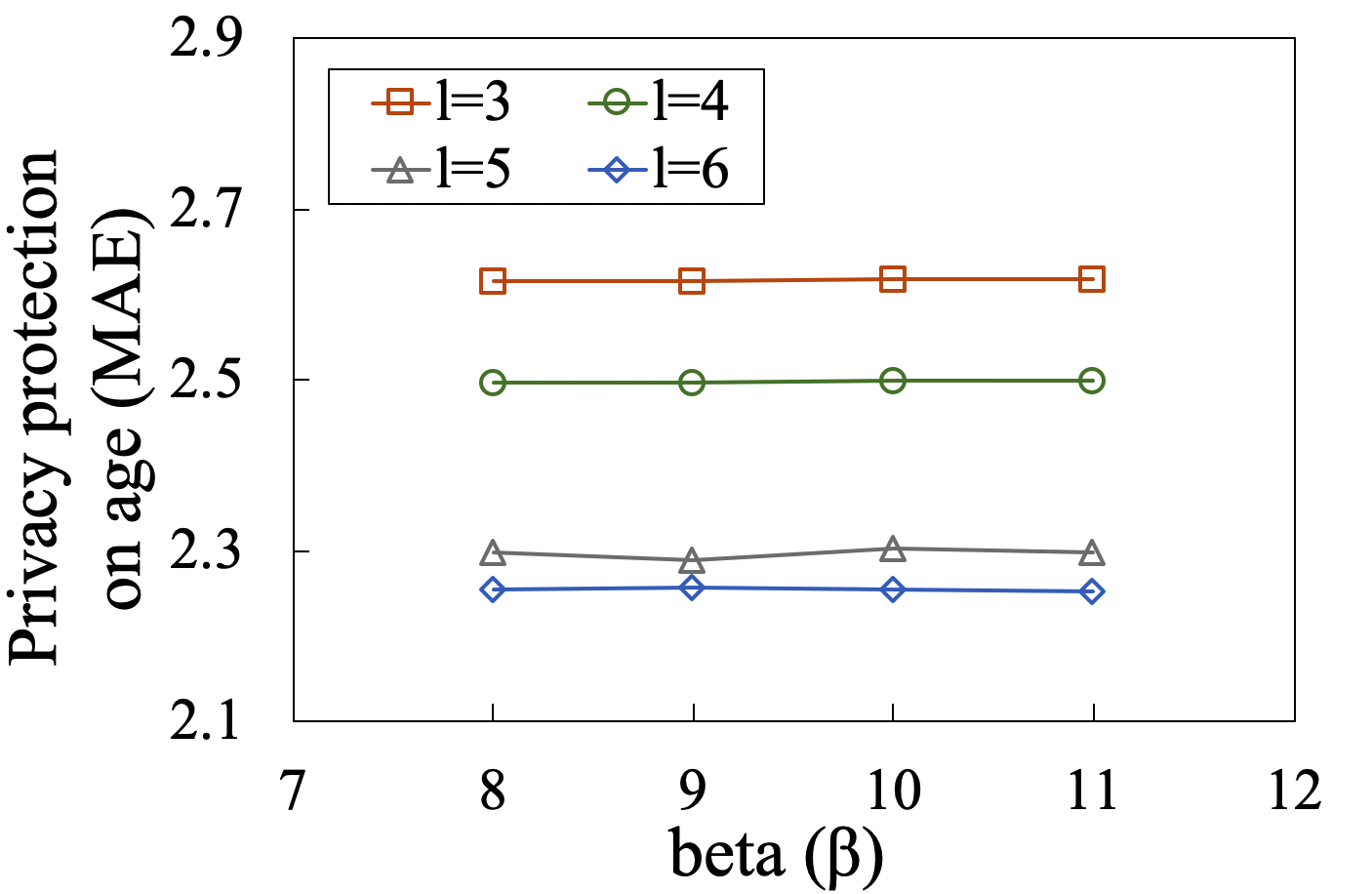}
		\caption{\emph{Age} (MovieLens)}
		\label{fig:ageLBeta}
	\end{subfigure}\ \ 
	\begin{subfigure}[b]{0.47\linewidth}
		\includegraphics[width=\textwidth]{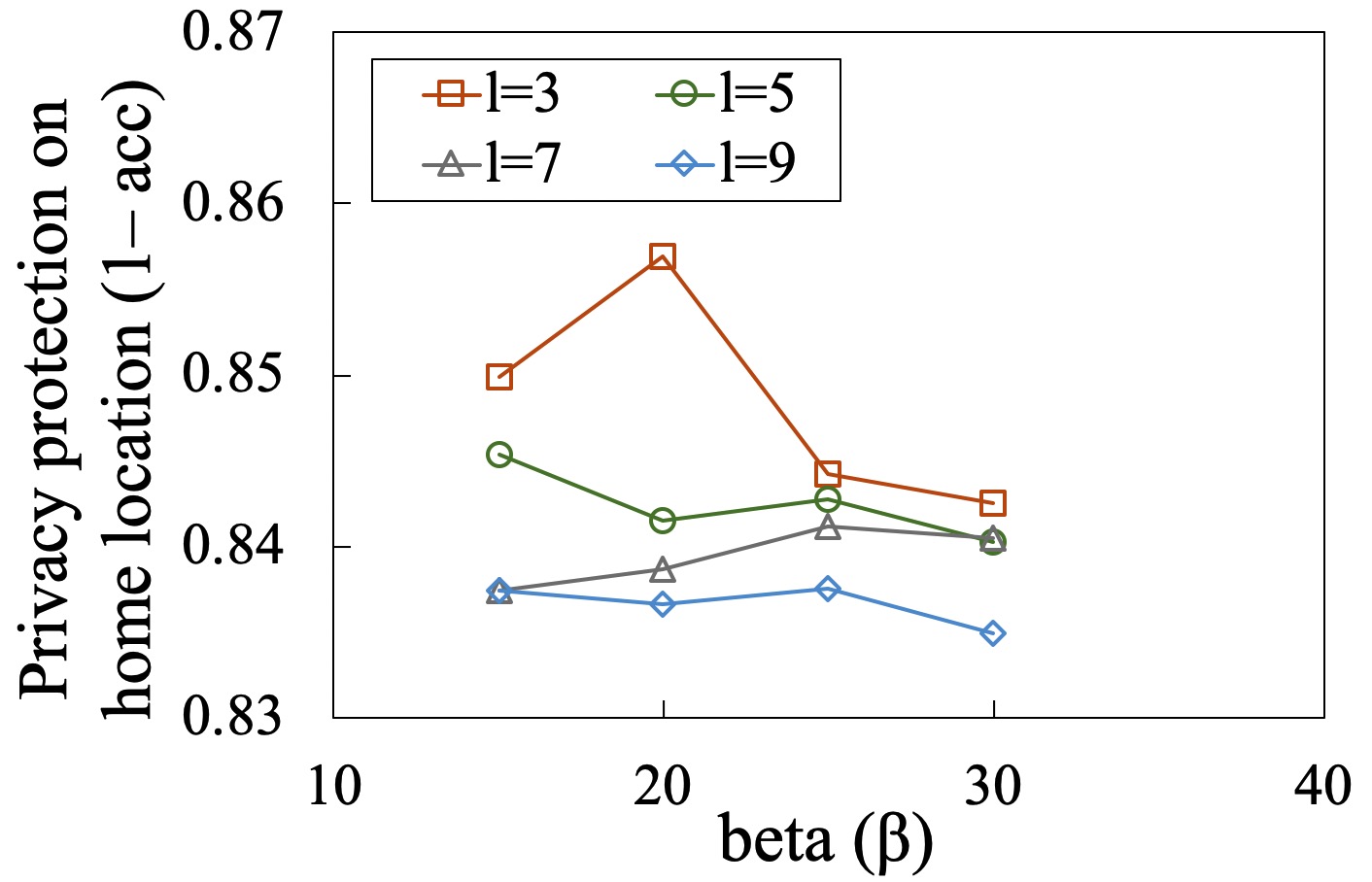}
		\caption{\emph{Location} (Foursquare)}
		\label{fig:locationLBeta}
	\end{subfigure}
	\caption{Protection by $(\boldsymbol{l},\boldsymbol{\beta})$-variety (Scenario~I, RF).}\label{fig:lbeta}
	\vspace{-0.5em}
\end{figure}

\begin{figure}[t!]
	\begin{minipage}{0.47\linewidth}
		\begin{center}
			\includegraphics[width=\linewidth]{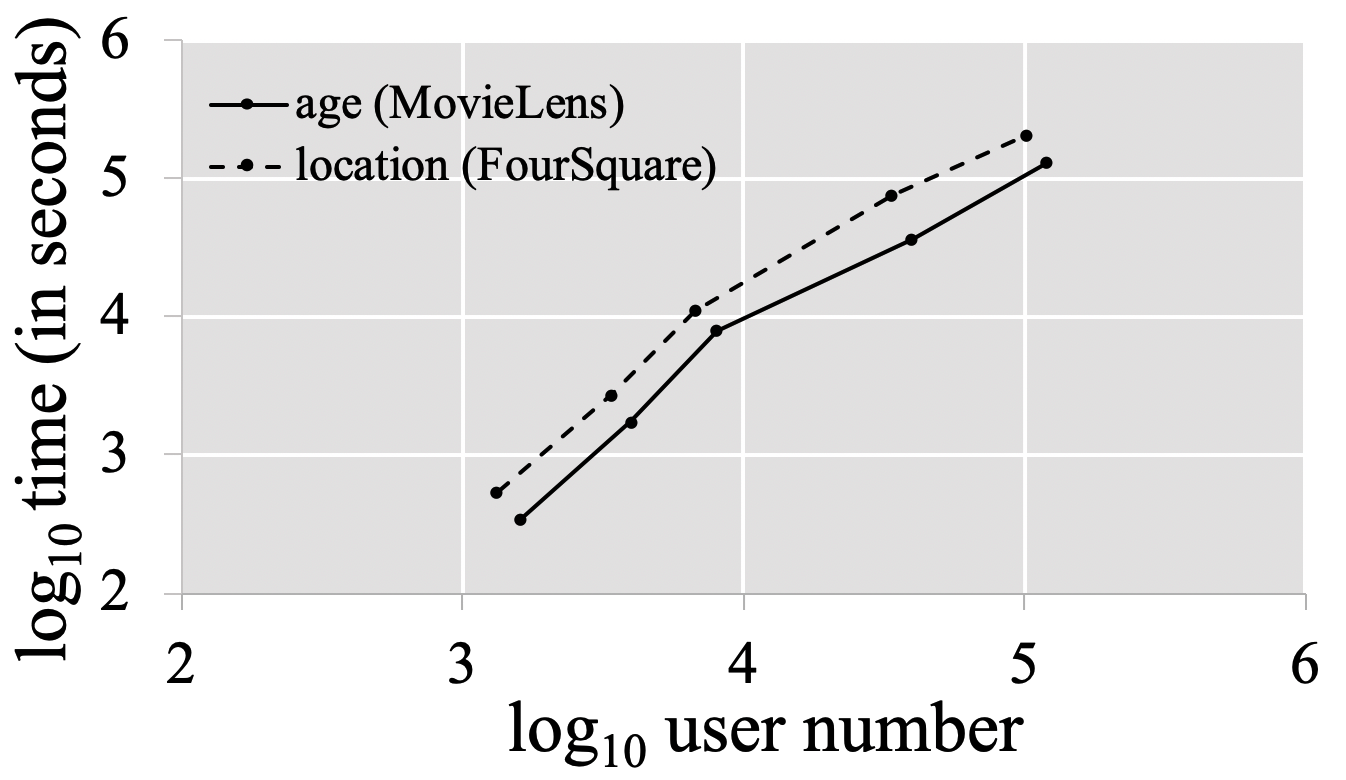}
			\caption{Scalability.}\label{fig:scalability}
			\vspace{-1em}
		\end{center}
	\end{minipage}\ \ 
	\begin{minipage}{0.47\linewidth}
		\begin{center}
			\includegraphics[width=\linewidth]{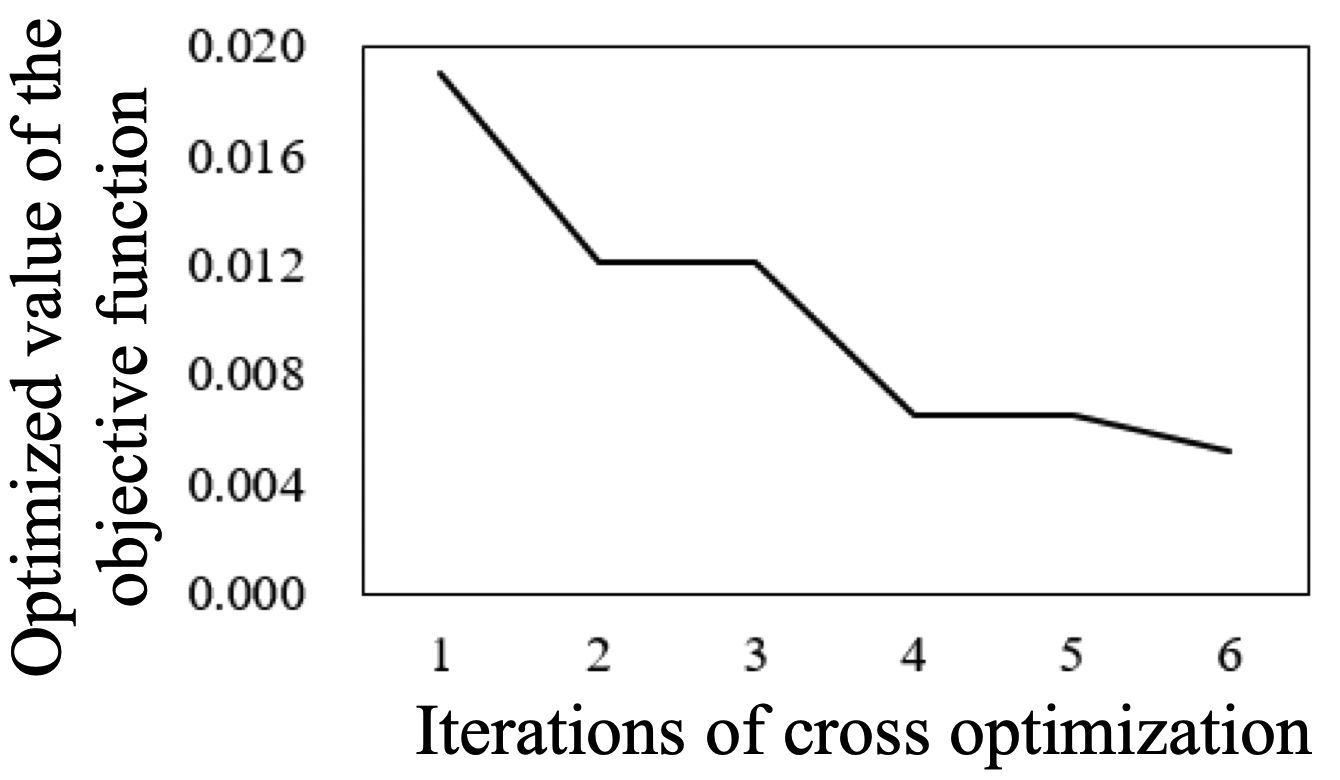}
			\caption{Convergence.}\label{fig:converge}
			\vspace{-1em}
		\end{center}
	\end{minipage}
\end{figure}

\textbf{The number of user clusters ($|\boldsymbol{C}|$)}.~As aforementioned, HyObscure and PrivCheck reduce computational complexity by clustering users based on their privacy-insensitive data. 
Here, we vary the number of clusters to compare the performances of HyObscure and PrivCheck. Fig.~\ref{fig:CN} reveals that HyObscure can consistently outperform PrivCheck when the number of user clusters varies. It is also worth noting that the privacy protection performance varies with number of user clusters for both HyObscure and PrivCheck. As a result, a good cluster number should be pre-searched to achieve a satisfactory privacy-utility tradeoff in reality.


\textbf{$(\boldsymbol{k},\boldsymbol{\alpha})$-uniqueness and $(\boldsymbol{l},\boldsymbol{\beta})$-variety}. 
Fig.~\ref{fig:kalpha} and Fig.~\ref{fig:lbeta} display how the privacy protection performance of HyObscure changes with $(k,\alpha)$ and $(l,\beta)$. Generally speaking, HyObscure performs very stably with varied parameters, and the greater impact comes from the parameters $k$ and $l$ rather than $\alpha$ and $\beta$. Note that a larger $k$ or $l$ given fixed $\alpha$ or $\beta$ indicates less utility of the generalized private data, and hence the privacy should be better protected by intuition. However, the experimental results show that the MAE of attackers' prediction actually drops, \ie, the privacy protection diminishes, when $k$ or $l$ increases. In other words, simply reducing data utility by increasing $k$ and $l$ may not lead to better protection against machine learning based inference attacks. One possible explanation is that we have more room to optimize the obscuring functions to achieve better protection when $k$ and $l$ are smaller with fewer restrictions.

\subsubsection{Algorithm Scalability}~We examine the computational efficiency of HyObscure by increasing the user size from 1000 to 100,000. Fig.~\ref{fig:scalability} shows that HyObscure scales linearly with user size on both data sets. Specifically, with our ordinary experimental environment, HyObscure is able to generate a privacy-preserving publishable dataset with around 100,000 users within 35 hours. Note that the obscuring functions are typically computed offline in practice. Thus, this computational efficiency could meet most requirements of moderate-scale data publishing, and can be sped up by using extra high-performance servers.

\subsubsection{Convergence of HyObscure}~While we have proved the convergence of HyObscure in theory in Sect.~\ref{sec:DataObscuring}, Fig.~\ref{fig:converge} empirically examines the convergence of the optimization process of an instance. It shows that the minimized privacy leakage decreases monotonically by the increase of optimization iterations with an apparent convergent trend. HyObscure stops when it cannot find a better solution to further reduce the objective value by more than the pre-defined threshold, which is set to 0.0001 in the experiment.

\section{RELATED WORK}
\label{sec:relate}
Traditional privacy-preserving data publishing is mainly studied in the database community, which often generalizes the fine-grained attribute value to the coarse one before data publishing, \ie, generalization techniques \cite{xu2014a}. $K$-anonymity \cite{Sweeney2002kAnonymityAM} and $L$-diversity \cite{Machanavajjhala2006lDiversityPB} are typical privacy metrics for quantifying the generalization protection effect. In particular, $K$-anonymity and $L$-diversity respectively ensure the number of users  in an equivalent class (i.e., with the same generalized value) larger than $K$ and the number of distinct private attribute values in an equivalent class larger than $L$~\cite{Sweeney2002kAnonymityAM,LeFevre2005IncognitoEF,LeFevre2006MondrianMK}. Larger $K$ and $L$ indicate better protection and less data utility.
Inspired by the definitions of $K$-anonymity and $L$-diversity, we propose $(k,\alpha)$-uniqueness and $(l,\beta)$-variety to restrict data utility loss and privacy leakage when generalizing users' private data (\eg, age).

The recent breakthrough in machine learning techniques brings many astonishing challenges on information protection \cite{yang2019privacy,salamatian2015managing}.  
The classical machine learning task needs original real data to learn the patterns in data which may cause data disclosure to attackers. Then, state-of-the-art work studies privacy-preserving machine learning schemes by allowing data patterns to be jointly learned from encrypted data without the need to share any real information~\cite{Mohassel2017SecureMLAS}. Moreover, a variety of machine learning based malicious attacks, such as attribute inference attack~\cite{gong2016you,Jia2017AttriInferIU} and membership inference attack~\cite{Shokri2016MembershipIA,Nasr2018MachineLW,Jia2019MemGuardDA}, appear and may lead to private data leakage or data membership disclosure through the published data and the well-trained model. This work is one of those focusing on data publishing techniques against inference attacks.

Delicately obscuring data before publishing is a typical idea to preserve private data being inferred by the published data while ensuring a required utility. In particular, obfuscating data or adding noises to original data are common means to trade off utility and privacy \cite{Jia2017AttriInferIU,yang2019privacy,salamatian2015managing}. 
Compared to prior research usually using only one obscuring technique, our work is the pioneering study focusing on hybrid obscuring, which simultaneously optimizes obfuscation (on  privacy-insensitive data) and  generalization (on private data). 


Quantification on privacy leakage and utility loss is critical for privacy-preserving data obscuring. 
Information theoretic metrics (\eg, differential privacy \cite{Dwork2006DifferentialP,McSherry2007MechanismDV,Wang2018GeographicDP}, mutual information \cite{Calmon2012PrivacyAS}, Kullback-Leibler divergence~\cite{lin1991divergence}, conditional entropy \cite{Sankar2013UtilityPrivacyTI}) are often used to quantify the privacy leakage, while utility loss of a privacy-preserving published data set is often measured by distance metrics (\eg, \textit{Hamming} and $l_2$-\textit{norm}, Jensen-Shannon divergence~\cite{lin1991divergence}). 
We propose new privacy leakage and utility loss measurements to adapt to our hybrid obscuring context.

\section{Conclusions}
\label{sec:conclude}
In this paper, we addressed a novel privacy-preserving data publishing problem that aims to employ hybrid obscuring operations on heterogeneous features for effective privacy-preserving data publishing. To that end, we first proposed new privacy and utility quantification measurements for published data by considering the joint effect of generalization and obfuscation. We also proposed a method called HyObscure to cross-iteratively optimize the data generalization and obfuscation functions for the best privacy protection effect under the utility guarantee. Both theoretical and empirical studies validated the effectiveness of HyObscure. 

Our study has some limitations. HyObscure in theory can work for multiple private features by treating them as a multi-dimensional feature. For example, \textit{Location} in our experiments is a two-dimensional feature with both latitude and longitude information. However, the searching space of generalization functions increases exponentially with the dimension, so the computational cost will be extremely high for very high-dimensional data. We leave the solution as future work. Moreover, while we take the widely adopted obfuscation and generalization operations as an example for hybrid obscuring of heterogeneous features, other obscuring measures like adding noise~\cite{Jia2017AttriInferIU} could also be integrated into HyObscure to resist more complicated inference attacks or attackers with more prior knowledge. We also leave this direction to future work. 

\bibliographystyle{IEEEtran}
\bibliography{relatedwork}

\normalsize
\section*{Supplemental Document}

\subsection*{Proof of Theorem 1}
The mutual information between $[\hat{X}, \widetilde Y]$ and $Y$~\cite{Calmon2012PrivacyAS} can be calculated as:
{
\footnotesize
\begin{equation}\label{eq:MI_loss}
\begin{aligned}
&I(\hat{X}, \widetilde Y; Y)= \\ &\sum_{\hat{x},y,\tilde y 
}p_{\hat{X}\widetilde Y Y}([\hat{x}, \tilde y],y)\log \frac{p_{\hat{X} \widetilde Y Y}([\hat{x},\tilde y],y)}{p_{\hat{X} \widetilde Y}([\hat{x}, \tilde y])} - \sum_{y} p_Y(y)\log p_Y(y).
\end{aligned}
\end{equation}
}

The first term of $I(\hat{X}, \widetilde Y; Y)$ can be rewritten as:
{
\footnotesize
\begin{equation}\label{eq:mutualInfo}
\begin{aligned}
&
\sum_{\tilde y 
} p_{\widetilde Y}(\tilde y) \sum_{\hat{x},y
} p_{\hat{X}Y}(\hat{x},y|\tilde y)\log \frac{p_{\hat{X}Y}(\hat{x},y| \tilde y)}{p_{\hat{X}}(\hat{x}| \tilde y)}\\
 = & \sum_{\tilde y,y'} 
 \mathcal G_{\widetilde{Y}|Y}(\tilde y|y)  p_{Y}(y')
 \sum_{\hat{x}, y }
 p_{\hat{X}Y}(\hat{x},y|\tilde y)\log \frac{p_{\hat{X}Y}(\hat{x},y| \tilde y)}{p_{\hat{X}}(\hat{x}| \tilde y)}.
\end{aligned}
\end{equation}
}
To keep the value distribution of the obfuscated privacy-insensitive attribute $X$ invariant to its original distribution before obfuscation given each generalized private value $\tilde y \in \widetilde Y$, we make the obfuscation occur inside the set of users with same $\tilde y \in \widetilde Y$.
Then, the joint probability of $\hat X$ and $Y$ given $\tilde y$ can be written as:
{
\footnotesize
\begin{equation}\label{eq:phatxyyo}
p_{\hat XY}(\hat x,y|\tilde y) = \sum_{x}
\mathcal{O}_{\hat X|X}(\hat x|x, \tilde y) \ p_{XY}(x,y| \tilde y).
\end{equation}}
Accordingly, $p_{\hat X}(\hat x| \tilde y)$ in Eq.~\eqref{eq:mutualInfo} can be obtained by:
{
\footnotesize
\begin{equation}\label{eq:phatX}
p_{\hat X}(\hat x| \tilde y) = \sum_{x,y} 
\mathcal{O}_{\hat X|X}(\hat x|x, \tilde y) \  p_{XY}(x,y| \tilde y).
\end{equation}
}
Integrating Eqs.~\eqref{eq:mutualInfo}-\eqref{eq:phatX} into Eq.~\eqref{eq:MI_loss}, we have Theorem~1.


\subsection*{Proof of Theorem 2}
Given a generalization function, solving $O$-problem outputs an obfuscation function that minimizes $I(\hat X, \widetilde{Y};Y)$; given an obfuscation function, solving $G$-problem outputs a generalization function that further reduces $I(\hat X, \widetilde{Y};Y)$. So, by iteratively solving $O$-problem and $G$-problem, $I(\hat X, \widetilde{Y};Y)$ decreases monotonically. Denote $I_0$ as $I(\hat X, \widetilde{Y};Y)$ obtained by solving $O$-problem given the initialized generalization function, and $\delta$ as the small threshold value to detect convergence. Then, \textit{HyObscure} will be terminated within $\lceil I_0/\delta \rceil$ iterations.

\subsection*{Proof of Theorem 3}
{
\footnotesize
	\begin{align}
		& I(\hat{X}, \widetilde Y; Y)  \\ = & \sum_{			\hat{x}, \tilde y, y}
			p_{\hat{X}\widetilde Y Y}([\hat{x}, \tilde y],y)\log \frac{p_{\hat{X} \widetilde Y Y}([\hat{x},\tilde y],y)}{p_{\hat{X} \widetilde Y}([\hat{x}, \tilde y])} - \sum_{y}
		p_Y(y)\log p_Y(y) \\
		= & \sum_{\hat{x}, \tilde y, y}
		p_{\hat{X}\widetilde Y Y}([\hat{x}, \tilde y],y)\log {p_{\hat{X} \widetilde Y Y}([\hat{x},\tilde y],y)} \nonumber \\&-\sum_{\hat{x}, \tilde y, y}
		p_{\hat{X}\widetilde Y Y}([\hat{x}, \tilde y],y)\log {p_{\hat{X} \widetilde Y}([\hat{x}, \tilde y])} - \sum_{y}
		p_Y(y)\log p_Y(y) \\
		= & \sum_{\hat{x}, y}
		p_{\hat{X} Y}(\hat{x},y)\log {p_{\hat{X} Y}(\hat{x},y)} \nonumber \\&-\sum_{\hat{x}, \tilde y}
		p_{\hat{X}\widetilde Y}(\hat{x}, \tilde y)\log {p_{\hat{X} \widetilde Y}(\hat{x}, \tilde y)} - \sum_{y}
		p_Y(y)\log p_Y(y) \\
		= & \sum_{\hat{x}, y}
		p_{\hat{X} Y}(\hat{x},y)\log \frac{p_{\hat{X} Y}(\hat{x},y)}{p_{\hat{X}}(\hat{x})} - \sum_{y}
		p_Y(y)\log p_Y(y) \nonumber \\&-\sum_{\hat{x}, \tilde y}
		p_{\hat{X}\widetilde Y}(\hat{x}, \tilde y)\log {p_{\hat{X} \widetilde Y}(\hat{x}, \tilde y)} + \sum_{\hat{x}}
		p_{\hat{X}}(\hat{x})\log{p_{\hat{X}}(\hat{x})} \\
		= & I(\hat{X}; Y) -\sum_{\hat{x}, \tilde y}
		p_{\hat{X}\widetilde Y}(\hat{x}, \tilde y)\log {p_{\hat{X} \widetilde Y}(\hat{x}, \tilde y)} + \sum_{\hat{x}}
		p_{\hat{X}}(\hat{x})\log{p_{\hat{X}}(\hat{x})}.
	\end{align}
}
According to Jensen inequality,  when $\sum_{i=1}^m{k_i} = 1,~k_i \in [0,1]$ we have
{
\footnotesize
	\begin{equation}
		\log \frac{1}{m} \le \sum_{i=1}^m k_i \log k_i \le 0.
	\end{equation}}
	
	Hence,
{
\footnotesize
	\begin{equation}
		I(\hat{X}, \widetilde Y; Y) \le I(\hat{X}; Y) - \log \frac{1}{|\hat X||\widetilde Y|}.
	\end{equation}
	}

Let the outputs of HyObscure be $\mathcal{O}_{\hat C|C}$ and $\mathcal G_{\widetilde Y|Y}$. Since $\mathcal{O}_{\hat C|C}$ is the optimal obfuscation function given $\mathcal G_{\widetilde Y|Y}$, we have
{
\footnotesize
	\begin{align}
		I(\hat{X}, \widetilde Y; Y|\mathcal{O}_{\hat C|C},\mathcal G_{\widetilde Y|Y}) \le & I(\hat{X}, \widetilde Y; Y|\mathcal{O}^*_{\hat C|C},\mathcal G_{\widetilde Y|Y}) \\ \le & I(\hat{X}; Y|\mathcal{O}^*_{\hat C|C}) - \log \frac{1}{|\hat X||\widetilde Y|}.
	\end{align}
}


\end{document}